\documentclass[journal,twoside]{IEEEtran}

\newtheorem{theorem}{Theorem}
\newtheorem{conjecture}[theorem]{Conjecture}

\usepackage{amsmath,color,psfrag,subfig,amssymb,epsfig,mathrsfs,cite}

\newcommand{\bb}[1]{{\color{blue} #1}}

\newcommand{\rect}{\mathop{\mathrm{rect}}\nolimits}

\makeatletter
\newcommand{\rmnum}[1]{\romannumeral #1}
\newcommand{\Rmnum}[1]{\expandafter\@slowromancap\romannumeral #1@}
\makeatother

\begin{document}
%
\title{Optimizing Constellations for Single-Subcarrier Intensity-Modulated Optical Systems}

%
%

\author{
Johnny Karout,
~Erik Agrell, 
Krzysztof Szczerba, and
Magnus Karlsson
\thanks{Manuscript received -; revised -. This work was supported by SSF under grant RE07-0026. The material in this paper was presented in part at the IEEE Global Communications Conference, Houston, TX, Dec. 2011.}
\thanks{J. Karout and E. Agrell are with the Department of Signals and Systems, Chalmers University of Technology, SE-412 96 Gothenburg, Sweden e-mail: ({johnny.karout, agrell}@chalmers.se).}
\thanks{K. Szczerba and M. Karlsson are with the Department of Microtechnology and Nanoscience, Chalmers University of Technology, SE-412 96 Gothenburg, Sweden e-mail: ({krzysztof.szczerba, magnus.karlsson}@chalmers.se).}
}

%
%

\markboth{IEEE TRANSACTIONS ON INFORMATION THEORY,~Vol.~-, No.~-, Month~Year}%
{Karout \MakeLowercase{\textit{et al.}}: Optimizing Constellations for Single-Subcarrier Intensity-Modulated Optical Systems}
%



\maketitle

\begin{abstract}

We optimize modulation formats for the additive white Gaussian noise channel with nonnegative input, also known as the intensity-modulated direct-detection channel, with and without confining them to a lattice structure. Our optimization criteria are the average electrical, average optical, and peak power.
The nonnegative constraint on the input to the channel is translated into a conical constraint in signal space, and modulation formats are designed by sphere packing inside this cone. 
Some dense packings are found, which yield more power-efficient modulation formats than previously known.
For example, at a spectral efficiency of 1.5 bit/s/Hz, the modulation format optimized for average electrical power has a 2.55 dB average electrical power gain over the best known format to achieve a symbol error rate of $10^{-6}$. 
The corresponding gains for formats optimized for average and peak optical power are 1.35 and 1.72 dB, respectively.
Using modulation formats optimized for peak power in average-power limited systems results in a smaller power penalty than when using formats optimized for average power in peak-power limited systems.
We also evaluate the modulation formats in terms of their mutual information to predict their performance in the presence of capacity-achieving error-correcting codes, and finally show numerically and analytically that the optimal modulation formats for reliable transmission in the wideband regime have only one nonzero point.

%
%


\bb{

}
\end{abstract}


%
\IEEEpeerreviewmaketitle

\section{Introduction }
\IEEEPARstart{C}{oherent} optical systems, which give access to both the carrier's amplitude and phase to convey information, allow the design of higher-order modulation formats which offer a good trade-off between spectral and power efficiency. 
%
However, in systems where phase information is absent, designing such formats becomes challenging.
Examples of such systems include phase-noise limited systems, and noncoherent systems where information is encoded onto the amplitude of the carrier and the envelope of the received signal is detected at the receiver. The latter is prevalent in optical communication systems where the overall cost and complexity is a critical constraint. This type of noncoherent systems is known as \emph{intensity-modulated direct-detection} (IM/DD) systems and will be the focus of our work. In such systems, the information is encoded onto the intensity of the optical carrier, and this intensity is, at all time instances, nonnegative.
Applications using IM/DD are, for example, wireless optical communications~\cite{Barry1994,Kahn1997,Hranilovic2004a} and short-haul fiber links used in, e.g., data centers~\cite{Randel2008,Molin2011}.

In the absence of optical amplification, an IM/DD system can be modeled as a conventional additive white Gaussian noise (AWGN) channel whose input is constrained to being nonnegative~\cite[Ch. 5]{Barry1994},~\cite{Kahn1997,Hranilovic2004,Hranilovic2003,Farid2010,Lapidoth2009}.
Since the optical phase cannot be used to carry information, resorting to $M$-ary pulse amplitude modulation ($M$-PAM) is a natural low-complexity way of improving the spectral efficiency beyond that of the widespread on-off keying (OOK).
However, this is different from the conventional PAM since no negative amplitudes can be used~\cite[Eq.~(5.8)]{Barry1994}. In~\cite{Cunningham2006,Szczerba2011opexpam}, an IM/DD link analysis using $4$-PAM signaling was demonstrated. In~\cite{Hranilovic2004}, upper and lower bounds on the capacity of $2$-, $4$-, $8$-, and $16$-PAM were derived and in~\cite{Walklin}, the power efficiency of $M$-PAM was shown to be low. The $M$-ary pulse-position modulation ($M$-PPM) formats are known to be power-efficient; however, they suffer from poor spectral efficiency~\cite[Sec. 5.3.3]{Barry1994},~\cite{Hranilovic2005,Kahn1997}.

Any nonnegative electrical waveform can be communicated successfully over an IM/DD link. This implies that if the information to be transmitted is modulated on an electrical subcarrier using any $M$-level modulation format, it can be transmitted on an IM/DD link after adding a direct current (DC) bias to ensure its nonnegativity, i.e., the subcarrier amplitude and phase which carry the information can be retrieved at the receiver. This concept is known as \emph{subcarrier modulation} (SCM) and was described in the wireless infrared communications context~\cite[Ch. 5]{Barry1994}.
Therefore, the power efficiency compared to $M$-PAM can be improved since SCM allows the use of power-efficient higher-order modulation  formats with IM/DD systems.
In~\cite{Wiberg2009}, the SCM concept was experimentally demonstrated, and in~\cite{Olsson2008a} and~\cite{Space}, a novel transmitter design for subcarrier quadrature phase-shift keying (QPSK) and 16-ary quadrature amplitude modulation (16-QAM) was presented.
As for the conventional electrical channel, many subcarriers can be superimposed resulting in a frequency division-multiplexing (FDM) system, referred to as multiple-subcarrier modulation (MSM) in the wireless infrared context~\cite[p. 122]{Barry1994}, and orthogonal frequency-division multiplexing (OFDM) if the carriers are orthogonal~\cite{Armstrong2009}. 
Further, a subclass of OFDM known as discrete multitone (DMT), where the output of the inverse fast Fourier transform modulator is real instead of complex, was investigated in~\cite{Lee2008}. 
In~\cite[Sec.~5.3.2]{Barry1994} and~\cite{Carruthers1996}, MSM was shown to have poor power efficiency compared to single-subcarrier modulation. Specifically, Barry~\cite[Sec.~5.3.2]{Barry1994} showed that the bandwidth of a multiple-subcarrier system is independent of the number of subcarriers, and that the average optical power penalty for multiple-subcarrier $M$-QAM in comparison with the single-subcarrier case is $5 \log_{10}N$ dB, where $N$ is the number of subcarriers. %
In~\cite{Randel2010}, DMT was shown to suffer considerably in peak-power limited systems. 
Kang and Hranilovic~\cite{Kang2006} and You and Kahn~\cite{You2001} proposed techniques to reduce the MSM power penalty.

For single-subcarrier modulation formats, one option is that the DC bias required to ensure the nonnegativity of the electrical waveform does not carry information~\cite[Ch. 5]{Barry1994},~\cite{Wiberg2009,Space}. The second option is by allowing the DC bias to carry information, thus potentially improving the power efficiency. 
This was studied by varying the DC bias on a symbol-by-symbol basis in~\cite{Hranilovic1999} and within the symbol interval in~\cite{You2001}.
By guaranteeing nonnegativity, the investigation of \emph{lattice codes} for IM/DD with AWGN became feasible~\cite{Shiu1999,Hranilovic2003}. 
Lattice codes, which are finite sets of points selected out of an $N$-dimensional lattice, have been extensively used in the construction of higher-order modulation formats for AWGN channels with coherent detection~\cite{Forney1988,Forney1989,Conway1999}. 
In addition, techniques such as constellation shaping and nonequiprobable signaling have been used to further minimize the average power~\cite{Forney1989,Calderbank1990}. 
In~\cite{Hranilovic2003}, a signal space model for optical IM/DD channels was presented, where average and peak optical power were considered as design constraints for constructing lattice-based modulation formats. Moreover, constellation shaping to reduce the average optical power were studied in~\cite{Shiu1999} for the case where no amplification is used, and in~\cite{Mao2008} where optical amplifiers are used. 
An interesting question is if there exist new constellations that perform better than already known ones, with or without coding. 

In this work, we address this question by optimizing single-subcarrier modulation formats for uncoded IM/DD systems, with and without confining them to a lattice structure. 
We choose a uniform probability distribution over constellation points as in~\cite{Barry1994,Kahn1997,Hranilovic2004a,Hranilovic2003,Olsson2008a,Space,Hranilovic1999,Karout2010,Szczerba2011,Essiambre2010}.
We propose a set of 4-, 8-, and 16-level single-subcarrier modulation formats which are optimized for average electrical, average optical, and peak power. These optimization criteria are all relevant, because the \emph{average electrical} power is the standard power measure in digital and wireless communications~\cite[p. 40]{Simon1995} and it helps in assessing the power consumption in optical communications~\cite{Chen1996}, while the \emph{average optical} power is an important figure of merit for skin- and eye-safety measures in wireless optical links~\cite[Ch. 5]{Barry1994},~\cite{Kahn1997,Hranilovic2003} and helps in quantifying the impact of shot noise in fiber-optical communications~\cite[p. 20]{Cox2002}. In addition, the \emph{peak} power, whether electrical or optical, is relevant for investigations of tolerance against the nonlinearities present in the system~\cite{Inan2009}. 
We then analyze the performance of the obtained modulation formats in terms of mutual information at different signal-to-noise ratios (SNR), in order to predict their performance in the presence of capacity-achieving error-correcting codes. 
%
Finally, we optimize modulation formats analytically in the wideband regime, i.e., at low SNR, while assuming uniform probability distributions and compare them with other formats.

The remainder of this paper is organized as follows. Section~\ref{sec:sysmodel} presents the system model. Section~\ref{sec:signalspace} elaborates on the signal space model, the performance measures, and the single-subcarrier modulation family which is the focus of this work. Section~\ref{sec:const} explains the optimization criteria that were used and describes the obtained modulation formats. In Section~\ref{sec:perf}, we evaluate the performance of these modulation formats in the absence and presence of capacity-achieving error-correcting codes and we compare them with already known modulation formats. 
In addition, we analytically optimize modulation formats for the low-SNR regime. In Section~\ref{sec:conclusion}, we summarize the main results and conclusions of this work.

\section{System Model}\label{sec:sysmodel}

\begin{figure}

\psfrag{1}[c][c][1][0]{$u(k)$ }
\psfrag{2}[c][c][1][0]{Modulator }
\psfrag{3}[c][c][1][0]{$x(t) \geq 0$ }
\psfrag{4}[c][c][1][0]{Laser diode }
\psfrag{5}[c][c][1][0]{$z(t)$ }
\psfrag{6}[c][c][1][0]{Optical link }
\psfrag{7}[c][c][1][0]{Photodetector }
\psfrag{8}[c][c][1][0]{$y(t) $ }
\psfrag{9}[c][c][1][0]{Demodulator }
\psfrag{a}[c][c][1][0]{$\hat u(k)$ }
\psfrag{b}[c][c][1][0]{$n(t)$ }
  \centering

  \subfloat[][]{\label{basebandtrans}\resizebox{3in}{!}{\includegraphics{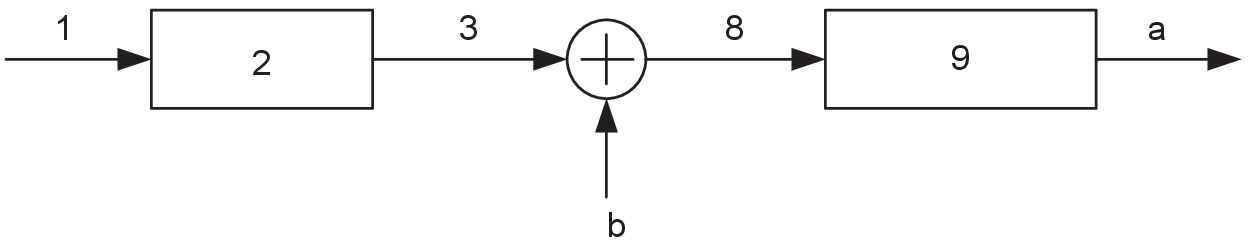}}}
  \hspace{3pt}
  \subfloat[][]{\label{Passbandtrans}
  \psfrag{1}[c][c][1.2][0]{$u(k)$ }
\psfrag{2}[c][c][1.2][0]{Modulator }
\psfrag{3}[c][c][1.2][0]{$x(t) \geq 0$ }
\psfrag{4}[c][c][1.2][0]{Light Source }
\psfrag{5}[c][c][1.2][0]{$z(t)$ }
\psfrag{6}[c][c][1.2][0]{Optical link }
\psfrag{7}[c][c][1.2][0]{Photodetector }
\psfrag{8}[c][c][1.2][0]{$y(t) $ }
\psfrag{9}[c][c][1.2][0]{Demodulator }
\psfrag{a}[c][c][1.2][0]{$\hat u(k)$ }

  \resizebox{3.4in}{!}{\includegraphics{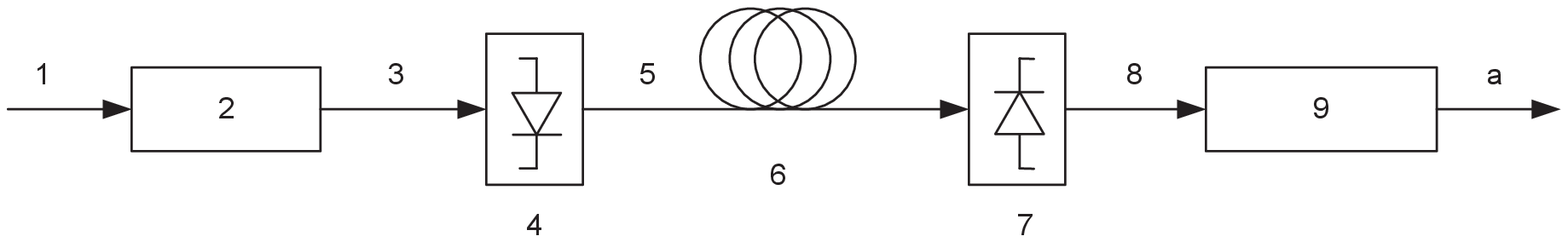}}}

  \caption{(a) Baseband transceiver with constrained-input Gaussian channel. (b) Passband transceiver of IM/DD systems.}
  \label{fig:model}

\end{figure}


The system model under study is depicted in Fig.~\ref{fig:model}(a). It consists of a modulator which maps the symbol $u(k)$ at instant $k$ to a waveform belonging to the signaling set $S=\{s_0(t),s_1(t), \ldots, s_{M-1}(t)\}$, where $T_s$ is the symbol period, $M$ is the size of the signaling set, and $s_i(t)=0$ for $t \notin [0,T_s)$ where $i=0,1,\ldots,M-1$. 
For the time-disjoint signaling case, the generated waveform
\begin{equation}\label{sig:xt}
x(t)=\sum_{k=-\infty}^{\infty} s_{u(k)}(t-k T_s),
\end{equation}
where $u(k)$
is an ergodic process uniformly distributed over $\{0,1,\ldots,M-1\}$, is constrained to being real and nonnegative.
The received signal can be written as
\begin{equation}\label{basebandreceivedsignal}
    y(t)= x(t) + n(t),
\end{equation}
where $n(t)$ is a zero-mean Gaussian process with double-sided power spectral density $N_0/2$. It should be noted that there exists no nonnegativity constraint on the signal $y(t)$. This is then followed by the demodulation of $y(t)$ which yields $\hat u(k)$, an estimate of $u(k)$. The demodulator is a correlator or matched filter receiver, which minimizes the symbol error rate at a given SNR~\cite[Sec. 4.1]{Simon1995}. This model is different from the conventional AWGN channel by the fact that the input $x(t)$ is constrained to being nonnegative.

The baseband model in Fig.~\ref{fig:model}(a) has been extensively studied in the optical communications context, since it serves as a good model for IM/DD systems~\cite[Ch. 5]{Barry1994},~\cite{Kahn1997,Hranilovic2004,Hranilovic2003,Farid2010,Lapidoth2009},~\cite[Sec. 11.2.3]{Hobook2005}. 
The passband transceiver for IM/DD systems is depicted in Fig.~\ref{fig:model}(b). In such systems, the electrical nonnegative waveform $x(t)$ directly modulates a light source, such as a laser diode. Therefore, the information is carried on the envelope of the passband signal
$z(t)=\sqrt{2 c x(t)} \cos( 2 \pi f_o t +\theta(t))$, i.e., the intensity of the optical field, where $c$ represents the electro-optical conversion factor in watts per ampere (W/A)~\cite{Cox2002,Westbergh2009,Coldren1999}, $f_o$ is the optical carrier frequency, and $\theta(t)$ is a random phase, uniformly distributed in $[0,2\pi)$ and slowly varying with $t$.
It then propagates through the optical medium depicted as an optical fiber in Fig.~\ref{fig:model}(b), which could be a free-space optical link in other applications. At the receiver, the photodetector detects the power of $z(t)$.
Since the dominant channel impairment in optical IM/DD systems is the thermal noise resulting from the optical-to-electrical conversion~\cite{Mao2008},~\cite[p. 155]{Agrawal2005}, the received electrical signal can be written as
\begin{equation}\label{ytwithslopeefficiency}
    y(t)= r c x(t) + n(t),
\end{equation}
where $r$ is the responsivity of the opto-electrical converter in A/W. Without loss of generality, we set $rc=1$, which yields~(\ref{basebandreceivedsignal}). 

There exists another IM/DD model which is relevant when the dominating noise comes from optical amplifiers, and not the receiver~\cite[Sec. 11.2]{Hobook2005},~\cite{Goebel2011,Channels2005}. The noise in that model has a noncentral $\chi^2$-distribution and not a Gaussian distribution as in this work.

\section{Signal Space Model}\label{sec:signalspace}

By defining a set of orthonormal basis functions $\phi_k(t)$ for $k=1,2, \ldots,N$ and $N \leq M$ as in~\cite{Hranilovic2003}, each of the signals in $S$ can be represented as
\begin{equation}\label{eachsignal}
    s_i(t)= \sum_{k=1}^{N} s_{i,k} \phi_k(t)
\end{equation}
for $i=0, \ldots, M-1$, where $\mathbf{s}_i=(s_{i,1},s_{i,2}, \ldots, s_{i,N})$ is the vector representation of $s_i(t)$ with respect to the aforementioned basis functions.
Therefore, the constellation representing the signaling set $S$ can be written as $\Omega=\{\mathbf{s}_0,\mathbf{s}_1, \ldots, \mathbf{s}_{M-1}\}$.
With this representation, the continuous-time channel models in~(\ref{basebandreceivedsignal}) and~(\ref{ytwithslopeefficiency}) can be represented by the discrete-time vector model
\begin{equation}\label{vecchannel}
    \mathbf{y}(k)= \mathbf{x}(k)+ \mathbf{n}(k),
\end{equation}
where, at instant $k$, $\mathbf{x}(k)\in \Omega$ is the transmitted vector and $\mathbf{n}(k)$ is a Gaussian random vector with independent elements, zero mean, and variance $N_0/2$ per dimension. Since $\mathbf{x}(k)$ and $\mathbf{y}(k)$ are both stationary processes, the argument $k$ will be dropped from now on.
To satisfy the nonnegativity constraint of the channel, the basis function $\phi_1(t)$ is set as in~\cite{Hranilovic2004,Hranilovic2003} to
\begin{equation}\label{basisfunction1}
  \phi_1(t) = \sqrt{ \frac{1}{T_s} } ~\rect\left(\frac{t}{T_s}\right),
\end{equation}
where
\begin{eqnarray*}
    \rect(t) =
\begin{cases}
 1, & \mbox{if } 0 \leq t \leq 1 \\
 0, & \mbox{otherwise.}
\end{cases}
\end{eqnarray*}
This basis function represents the DC bias. Thus, $s_{i,1}$ is chosen for each $i=0,\ldots,M-1$ such that
\[\min_{t} s_i(t) \geq 0,\] which guarantees the nonnegativity of $x(t)$ in~(\ref{sig:xt}).
The admissible region $\Upsilon$ containing the set of all signal vectors satisfying the nonnegativity constraint can be represented as~\cite[Eq.~(10)]{Hranilovic2003}
\begin{equation}\label{adm}
    \Upsilon= \{\mathbf{w} \in \mathbb{R}^N :\min_{t \in [0,T_s)} \sum_{k=1}^{N} w_{k} \phi_k(t) \geq 0\},
\end{equation}
where $\mathbf{w}=(w_{1},w_{2}, \ldots, w_{N})$. 
Therefore, the constellation $\Omega$ is a finite subset of $\Upsilon$.
The admissible region $\Upsilon$ for IM/DD systems has been shown in~\cite[Th.~1]{Hranilovic2003} to be the convex hull of a generalized $N$-dimensional cone with vertex at the origin and opening in the dimension spanned by $\phi_1(t)$.

\subsection{Performance Measures}
Unlike the conventional electrical AWGN channel where the two standard power performance measures are the average and peak electrical power, three important performance measures for IM/DD channels can be extracted from the baseband and passband models in Fig.~\ref{fig:model}. The first entity is the average electrical power defined as
\begin{equation*}\label{aveele}   
   \bar P_{e}= \lim_{T \to \infty}   \frac{1}{2T} \int_{-T}^{T}    \! x^2(t) \,  \mathrm{d}t,
\end{equation*}
which for any basis functions can be simplified to
\begin{equation}\label{aveelesig}
   \bar P_{e}=  \frac{{E_s}}{T_s}  =\frac{1}{T_s}~\mathbb{E}[ \|\mathbf{s}_I \|^2] ,
\end{equation}
where ${E_s}$ is the average energy of the constellation, $\mathbb{E}[\cdot]$ is the expected value, and $I$ is a random variable uniformly distributed over $\{ 0,1,\ldots,M-1\}$. This entity is an important figure of merit for assessing the performance of digital and wireless communication systems~\cite[p. 40]{Simon1995}. Therefore, it is relevant for IM/DD systems for compatibility with classical methods and results~\cite{Channels2005,SvalutoMoreolo2010}. In addition, it helps in quantifying the impact of relative intensity noise (RIN) in fiber-optical links~\cite{Cox2002}, and in assessing the power consumption of optical systems~\cite{Chen1996}. In~\cite{Karout2010}, $\bar P_{e}$ was used as a performance measure for comparing different intensity modulation formats.

The second measure is the average optical power $\bar P_o$, which has been studied in~\cite{Barry1994,Kahn1997,Hranilovic2004,Hranilovic2003,Farid2010} for the wireless optical channel. Limitations are set on $\bar P_o$ for skin- and eye-safety standards to be met. In fiber-optic communications, this entity is used to quantify the impact of shot noise on the performance~\cite[p. 20]{Cox2002}. It is defined as
 \begin{equation*}\label{opticalpower}
    \bar P_{o}=  \lim_{T\to \infty}\frac{1}{2T} \int_{-T}^{T} \! z^2(t) \, \mathrm{d}t= \lim_{T \to \infty} \frac{c}{2T} \int_{-T}^{T} \!  x(t) \,  \mathrm{d}t.
  \end{equation*}
This measure depends solely on the DC bias required to make the signals nonnegative and can be represented in terms of the symbol period and constellation geometry as~\cite{Hranilovic2004,Hranilovic2003}
\begin{equation}\label{opticalpowersig}
    \bar P_{o}=  \frac{c}{\sqrt{T_s}}~\mathbb{E} [s_{I,1}],
      \end{equation}
regardless of $\phi_2(t), \ldots, \phi_N(t)$.

The third measure is the peak optical power defined as
\begin{eqnarray}\label{peakoptpowersig}
 \nonumber  \hat{P}_o &=& \max_{t} \frac{z^2(t)}{2} \\ 
   &=&c \max_{t} x(t).
\end{eqnarray}
It is relevant for investigations of tolerance against the nonlinear behavior of transmitting and receiving hardware in communication systems~\cite{Inan2009,Hranilovic2003,Coldren1999} and has been studied in~\cite{Hranilovic2003,Farid2010,Karout2010}.
The peak electrical power $\hat P_e$ is directly related to $\hat P_o$ by \[\hat P_e =   \bigg( \frac{\hat P_o}{c }  \bigg)^2  ,\]
and will not be further considered in this paper, since a constellation optimized for $\hat P_o$ will automatically be optimized for $\hat P_e$ too. A general form for $\hat P_o$ as a function of $\Omega$, as in~(\ref{aveelesig}) and~(\ref{opticalpowersig}) for $\bar P_e$ and $\bar P_o$, does not exist, since $\hat P_o$ depends on the exact choice of basis functions. A special case will be studied in Sec.~\ref{sec:scm}.

To assess the performance of the different modulation formats in the presence of capacity-achieving error-correcting codes, we consider the mutual information~\cite[Sec. 2.4]{Cover2006} \begin{equation}\label{MIformula}
    I(\mathbf{x};\mathbf{y})= H(\mathbf{x}) -H(\mathbf{x}|\mathbf{y})
\end{equation}
as a performance measure.
The terms $H(\mathbf{x})$ and $H(\mathbf{x}|\mathbf{y})$ are the entropy of $\mathbf{x}$ and the conditional entropy of $\mathbf{x}$ given the received vector $\mathbf{y}$, averaged over both $\mathbf{x}$ and $\mathbf{y}$.
The channel capacity of a discrete memoryless channel is
\cite[Eq.~(7.1)]{Cover2006}
\begin{equation}
C= \max_{{p}(\mathbf{x})} I(\mathbf{x};\mathbf{y}),
\end{equation}
where the maximum is taken over all possible input distributions ${p}(\mathbf{x})$.
For a fixed constellation and distribution, the mutual information gives a lower bound on the channel capacity. In our work, like in the works of many other authors, we choose a uniform distribution over the constellation points.

%

We define $R_s=1/T_s$ as the symbol rate in symbols per second, $R_b=R_s R$ as the bit rate in bits per second, and $E_b=E_s /R$ as the average energy per bit. Furthermore, in order to have a fair comparison 
of the bit rates that can be achieved by the different modulation formats in a fixed bandwidth, the spectral efficiency defined as
\[
\eta=\frac{R_b}{W}
 ~\text{[bit/s/Hz]}
\]
should be taken into account, where $W$ is the baseband bandwidth defined as the first null in the spectrum of $x(t)$. 
In this paper, we are interested in two extreme cases: the uncoded system, for which $R=\log_2 M$, and the system with optimal coding, for which $R=  I(\mathbf{x};\mathbf{y})$.


\subsection{Single-Subcarrier Modulation Formats}\label{sec:scm}

For in-phase and quadrature phase (I/Q) modulation formats to be used on intensity modulated channels, a DC bias is required in order for $x(t)$ to be nonnegative. This could be translated geometrically by having a three-dimensional (3d) Euclidean space spanned by the orthonormal basis functions $\phi_1(t)$ defined in~(\ref{basisfunction1}) and
\begin{eqnarray}\label{basisfunctions}
 \label{phi2} \phi_2(t) &=& \sqrt{ \frac{2}{T_s} }~\cos{(2\pi f t)} ~\rect\left(\frac{t}{T_s}\right),\\
 \label{phi3} \phi_3(t) &=& \sqrt{ \frac{2}{T_s} }~\sin{(2\pi f t)}~\rect\left(\frac{t}{T_s}\right),
\end{eqnarray}
which are the basis functions of conventional I/Q modulation formats such as $M$-PSK and $M$-QAM, where $f$ is the electrical subcarrier frequency~\cite{Hranilovic2003,Barry1994}.
As in~\cite[pp. 115--116]{Barry1994} and~\cite{Hranilovic2003}, we use $f=1/T_s$, which is the minimum value for which $\phi_1(t)$, $\phi_2(t)$, and $\phi_3(t)$ are orthonormal. In~\cite{Hranilovic2003}, IM/DD modulation formats based on these three basis functions are referred to as raised-QAM, and in~\cite{Wiberg2009} as single cycle SCM.
At the same symbol rate, modulation formats such as OOK and $M$-PAM have $W=R_s$, whereas the modulation formats belonging to the single-subcarrier family occupy $W=2 R_s$; this is due to the intermediate step of modulating the information onto an electrical subcarrier before modulating the optical carrier~\cite[Ch. 5]{Barry1994},~\cite{Karout2010}.

We now describe explicitly the admissible region $\Upsilon$ for single-subcarrier modulation formats~\cite[Fig. 4.2]{Hranilovic2004a},~\cite{Hranilovic2003}.

\begin{theorem}\label{theorem:cone:admi}
For the specific set of basis functions $\phi_1(t)$, $\phi_2(t)$, and $\phi_3(t)$ defined in~(\ref{basisfunction1}), (\ref{phi2}), and~(\ref{phi3}), the admissible region $\Upsilon$ is a three-dimensional (3d) cone with vertex at the origin, apex angle of $\cos^{-1}(1/3)=70.528^{\circ}$, and opening in the dimension spanned by $\phi_1(t)$.
\end{theorem}

\begin{IEEEproof}
The admissible region in~(\ref{adm}) can be written for single-subcarrier modulation formats as
\begin{eqnarray} \label{admissibile_step1}
    \Upsilon & = & \{\mathbf{w} \in \mathbb{R}^3 :\min_{t \in [0,T_s)} \sum_{k=1}^{3} w_{k} \phi_k(t)\geq 0\  \},  
    \end{eqnarray}
    where
    \begin{eqnarray} 
   \nonumber & & \min_{t \in [0,T_s)} \sum_{k=1}^{3} w_{k} \phi_k(t)   \\ \nonumber
    & & = \frac{1}{\sqrt{T_s} }
     \min_{t \in [0,T_s)} \big\{
     w_{1}   \\ \nonumber
 & &~~ + \sqrt{2(w_{2}^2 + w_{3}^2)} \cos{\big(2\pi f t - \theta \big) } \big\} \\ 
 \label{lastmin}  & & =  \frac{1}{\sqrt{T_s} }  \big( w_{1}  - \sqrt{2(w_{2}^2 + w_{3}^2)} \big),
    \end{eqnarray} 
    where $\theta= \arg( w_{2} + j w_{3})$.
    Therefore, substituting (\ref{lastmin}) in (\ref{admissibile_step1}) yields
  \begin{eqnarray}
 \label{adm:scm}   \Upsilon  & = & \{\mathbf{w} \in \mathbb{R}^3 : w_{1} \geq \sqrt{2(w_{2}^2 +w_{3}^2)}\},
\end{eqnarray}
which  is a 3d-cone with apex angle of $\cos^{-1}(1/3)=70.528^{\circ}$ pointing in the dimension spanned by $\phi_1(t)$, with vertex at the origin.
\end{IEEEproof}

The average electrical and optical power were given in~(\ref{aveelesig}) and (\ref{opticalpowersig}) as functions of the constellation $\Omega$; however, the peak optical power defined in~(\ref{peakoptpowersig}) could for these basis functions be expressed in terms of the constellation geometry too~\cite[Fig. 4.3]{Hranilovic2004a}.

\begin{theorem}
The peak optical power for the single-subcarrier modulation formats with the above defined basis functions can be expressed as
\begin{equation}\label{peakpower}     
   \hat{P}_o =\frac{c}{\sqrt{T_s}}~ \max_{i} \left\{s_{i,1} +\sqrt{2(s_{i,2}^2+ s_{i,3}^2)} ~\right\}.
\end{equation}
\end{theorem}

\begin{IEEEproof}
From~(\ref{peakoptpowersig}), $ \hat{P}_o $ can be written as
\begin{eqnarray*}
\hat{P}_o  &=& c \max_{i,t} s_i(t)\\
   &=& c \max_{i,t} \sum_{k=1}^{3} s_{i,k} \phi_k(t) \\
   &=& \frac{c}{\sqrt{T_s}}~ \max_{i} \left\{s_{i,1} +\sqrt{2(s_{i,2}^2+ s_{i,3}^2)} ~\right\}.
\end{eqnarray*}
Alternatively, the theorem can be proved using~\cite[Th.~2]{Hranilovic2003}.
\end{IEEEproof}

%


%

\begin{figure}
\centering
\psfrag{a}[c][c][0.65][0]{$ \phi_1(t)$ }

\psfrag{b}[c][l][0.65][0]{$ \phi_2(t), \phi_3(t)$ }

\includegraphics[width=0.5\textwidth]{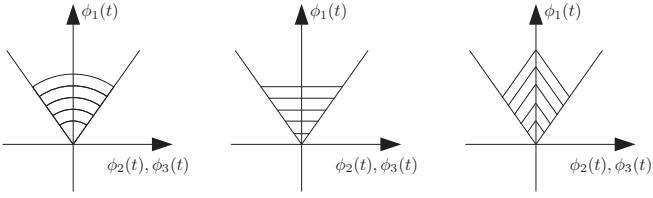}

\caption{(left to right): Contours of equal $\bar P_e$, $\bar P_o$, and $\hat P_o$. }
\label{contour}
\end{figure}

\section{Constellation Optimization}\label{sec:const}

To design power-efficient constellations, the admissible region in~(\ref{adm:scm}) has to be taken into account. As done before for the conventional AWGN channel~\cite{Foschini1974,Porath2003,Sloane1995,Graham1990,Agrell2009}, our approach of finding the best constellations can be formulated as a sphere-packing problem with the objective of minimizing a cost function depending on the constraints that might be present in the system model shown in Fig.~\ref{fig:model}. Thus, the optimization problem, for given constants $M$ and $d_{\text{min}}$, can be written as
\begin{eqnarray}\label{obj}
\label{minline} \text{Minimize~} & & \xi(\Omega)  \\
\label{minline1} \text{Subject to}& & |\Omega| =M\\
\label{cons2}   & & \Omega \subset \Upsilon\\
\label{cons1}   & &  d(\Omega) = d_{\text{min}},
\end{eqnarray}
where \[d(\Omega)=\min_{  \underset{i \ne j }{\mathbf{s}_i ,\mathbf{s}_j \in \Omega}} \| \mathbf{s}_i -\mathbf{s}_j \|.\]

Choosing the objective function as $\xi(\Omega)=\mathbb{E}[ \|\mathbf{s}_I \|^2]$ results in $\Omega=\mathscr{C}_{\bar P_e,M}$, i.e., a constellation optimized for average electrical power, and $\xi(\Omega)=\mathbb{E} [s_{I,1}]$ results in $\Omega=\mathscr{C}_{\bar P_o,M}$, i.e., a constellation optimized for average optical power. Finally,
\[ \xi(\Omega)= \max_{i} \left\{s_{i,1} +\sqrt{2(s_{i,2}^2+ s_{i,3}^2)} ~\right\}\]
yields $\Omega=\mathscr{C}_{\hat P_o,M}$, a constellation optimized for peak optical power.
Fig.~\ref{contour} depicts a two-dimensional contour plot of the three objective functions together with the admissible region $\Upsilon$.
The constraint in~(\ref{cons2}) guarantees that the signals belong to the admissible region $\Upsilon$, therefore satisfying the nonnegativity criterion of the channel.
%
%
The minimum distance $d_{\text{min}}$ in~(\ref{cons1}) serves as a good measure of error probability performance in the presence of AWGN at high SNR.
Although this optimization problem is well formulated mathematically, it is difficult to obtain an analytical solution. Therefore, we resorted to \emph{numerical optimization} techniques as in~\cite{Foschini1974,Porath2003,Sloane1995,Graham1990,Agrell2009} to find the best constellations. 
One drawback of such constellations is often the lack of geometric regularity, which increases the modulator and demodulator complexity. 
The optimization problem is nonconvex; therefore, a local solution does not imply that it is globally optimal.


A special case of this optimization problem, which might not guarantee the optimal solution, is to confine the possible constellations to have a regular structure such as that of a lattice, denoted by $\Lambda$. In this case, the above optimization problem can be reformulated by replacing~(\ref{cons2}) with  $ \Omega \subset \Upsilon \cap \Lambda$, and dropping~(\ref{cons1}) since it is directly inferred by $ \Omega \subset \Lambda$. 
This was done in~\cite{Hranilovic2003} for the cubic and Leech lattices. In order to compare with the best nonlattice constellations of relatively small sizes $M$, we use in this work
 the face-centered cubic lattice ($A_3$), which provides the densest packing for the 3d-Euclidean space~\cite[p.~\rmnum{16}  ]{Conway1999}. The obtained lattice-based constellations optimized for average electrical, average optical, and peak optical power are denoted $\mathscr{L}_{\bar P_e,M}$, $\mathscr{L}_{\bar P_o,M}$, and $\mathscr{L}_{\hat P_o,M}$, respectively.

\begin{figure*}
\centering
\begin{tabular}{c}

\includegraphics[width=0.23\textwidth]{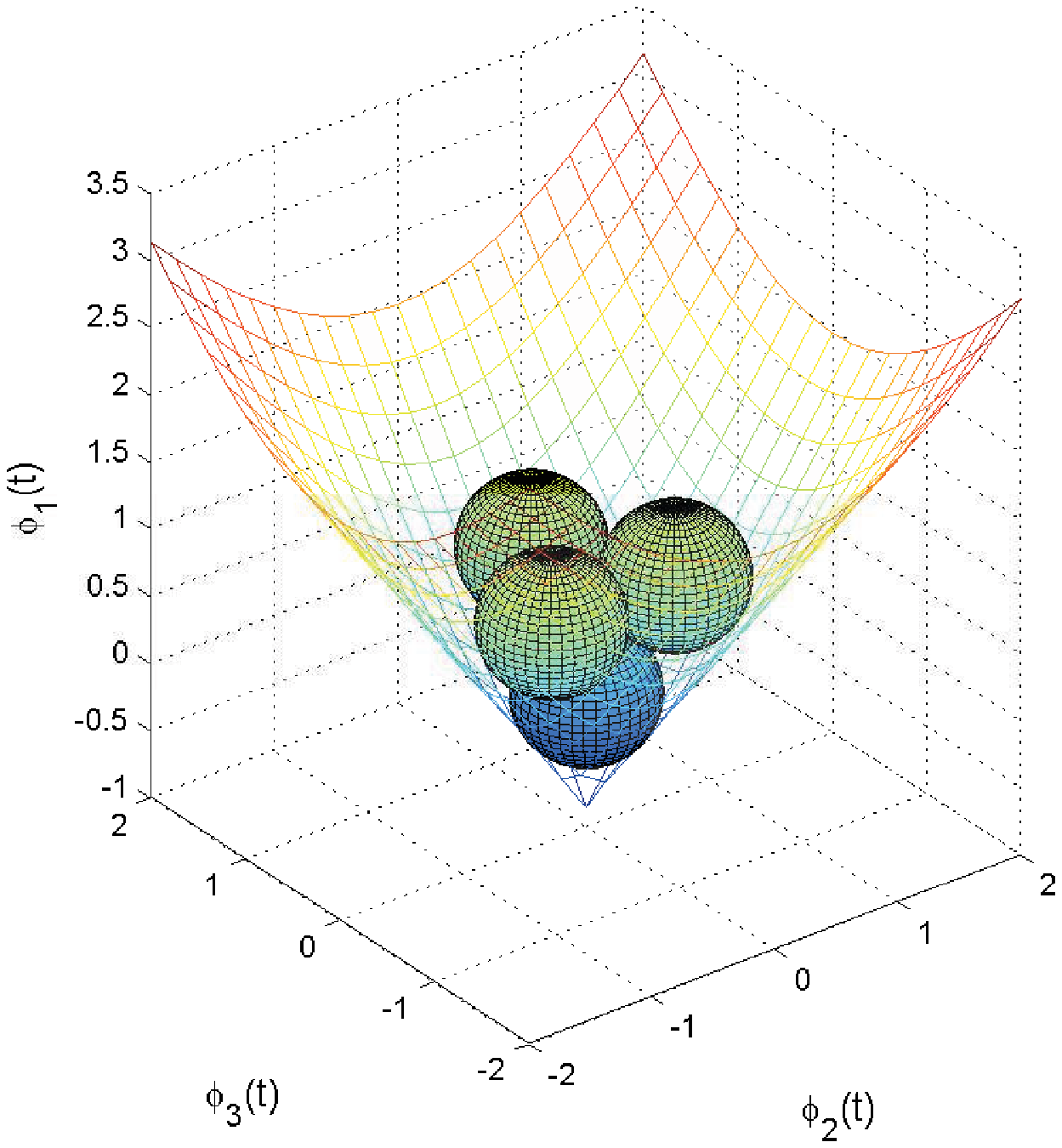}
\end{tabular}
\caption{$\mathscr{C}_{4}$ = $\mathscr{L}_{4}$. }
\label{4pointscone}

\begin{tabular}{cccc}
\includegraphics[width=0.23\textwidth]{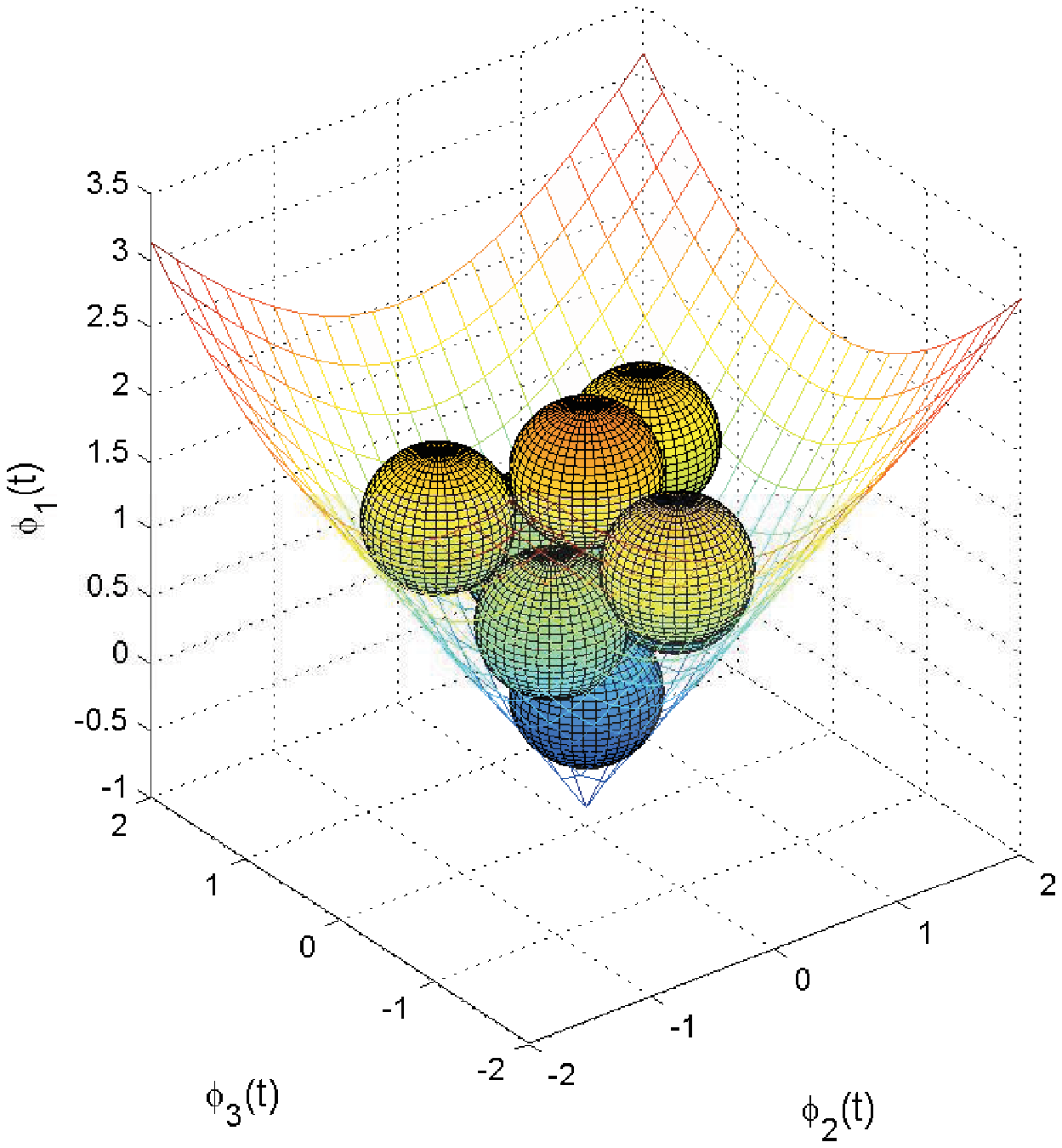} &
\includegraphics[width=0.23\textwidth]{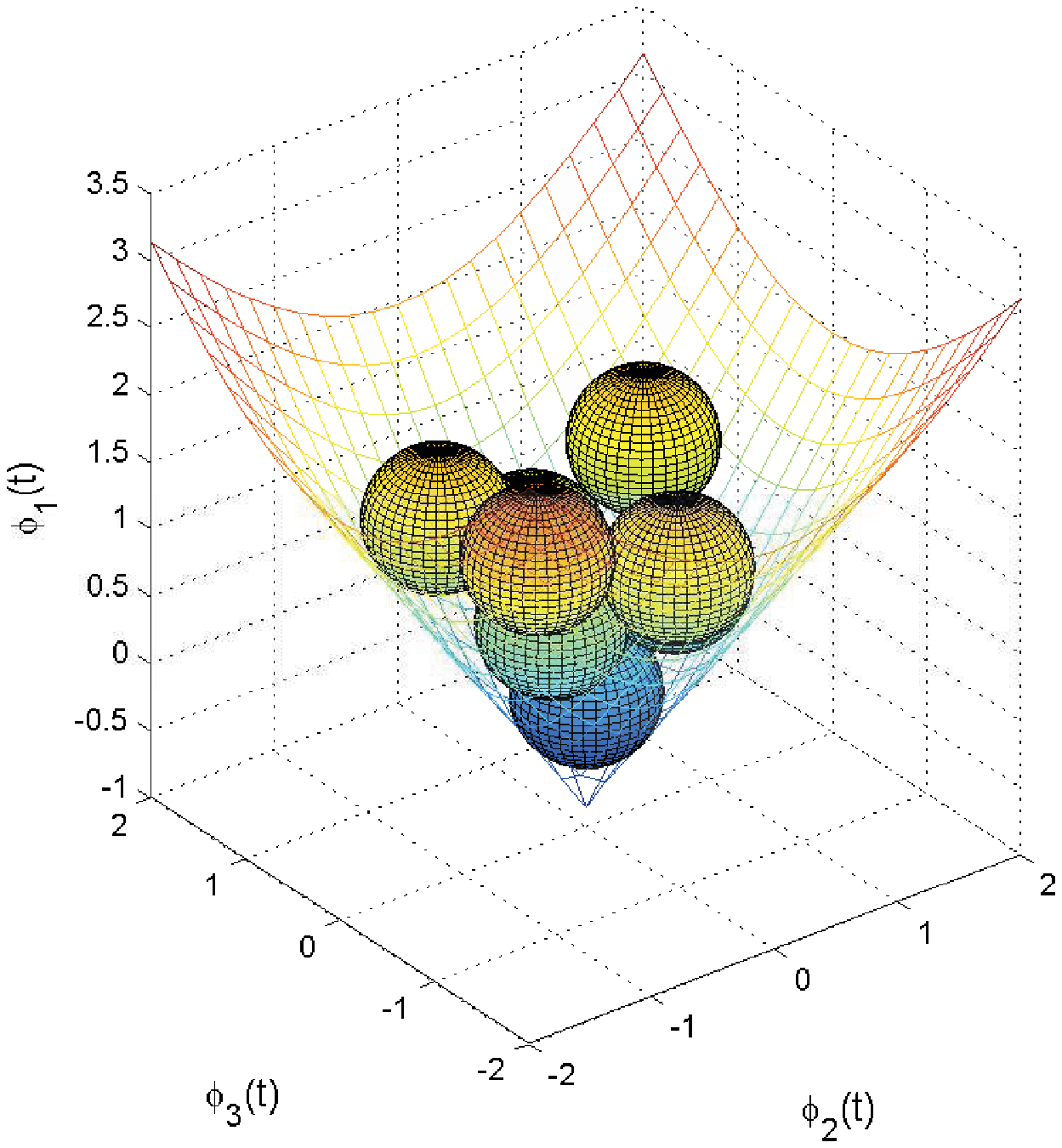} &
\includegraphics[width=0.23\textwidth]{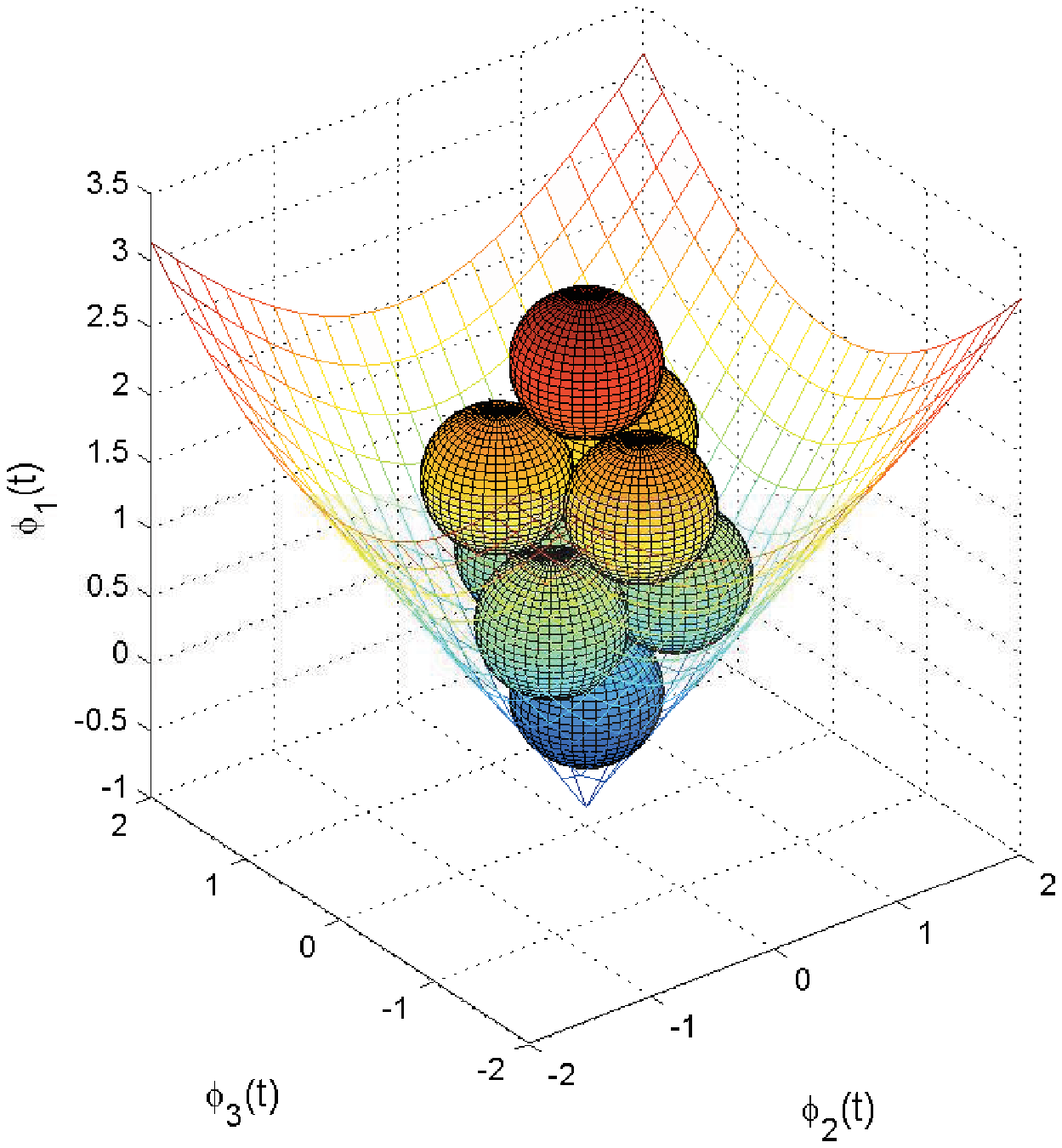} &
\includegraphics[width=0.23\textwidth]{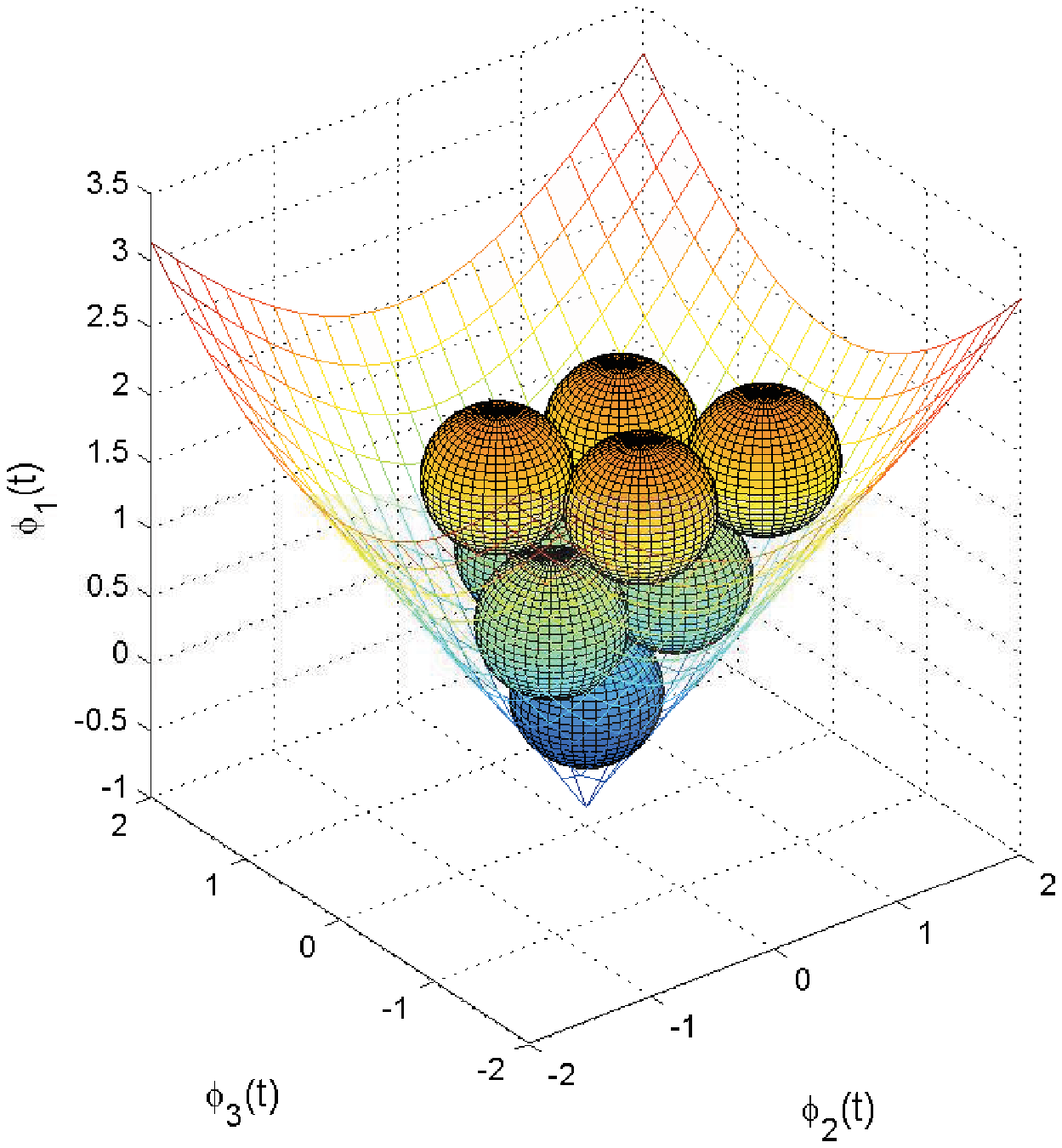}

\end{tabular}
\caption{(left to right): $\mathscr{C}_{\bar P_e,8}$,
 $\mathscr{C}_{\bar P_o,8}$,
   $\mathscr{C}_{\hat P_o,8}$ = $\mathscr{L}_{\hat P_o,8}$,
   $\mathscr{L}_{\bar P_e,8}$ = $\mathscr{L}_{\bar P_o,8}$.}
\label{8pointscone}

\begin{tabular}{cccc}
\includegraphics[width=0.23\textwidth]{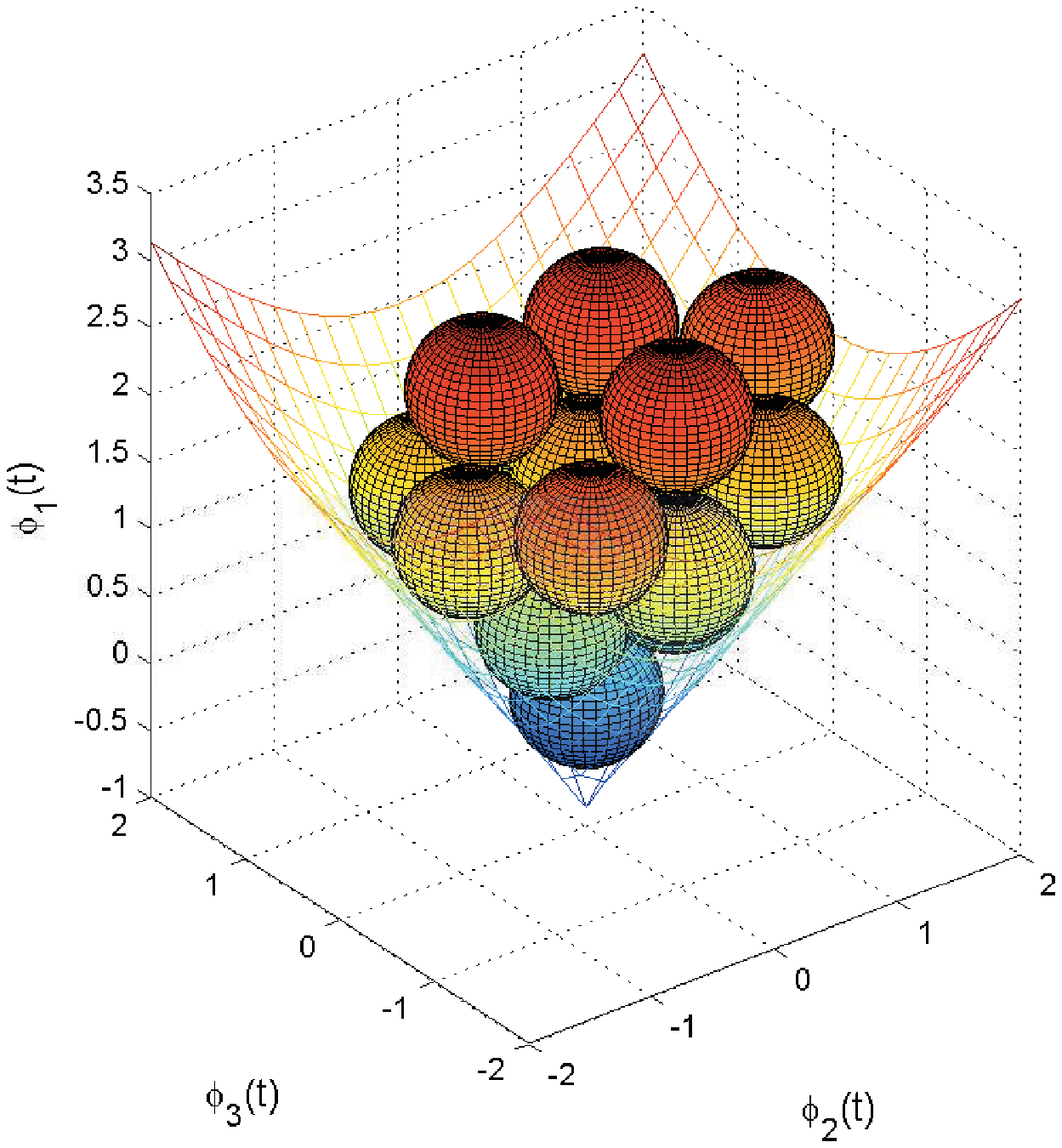} &
\includegraphics[width=0.23\textwidth]{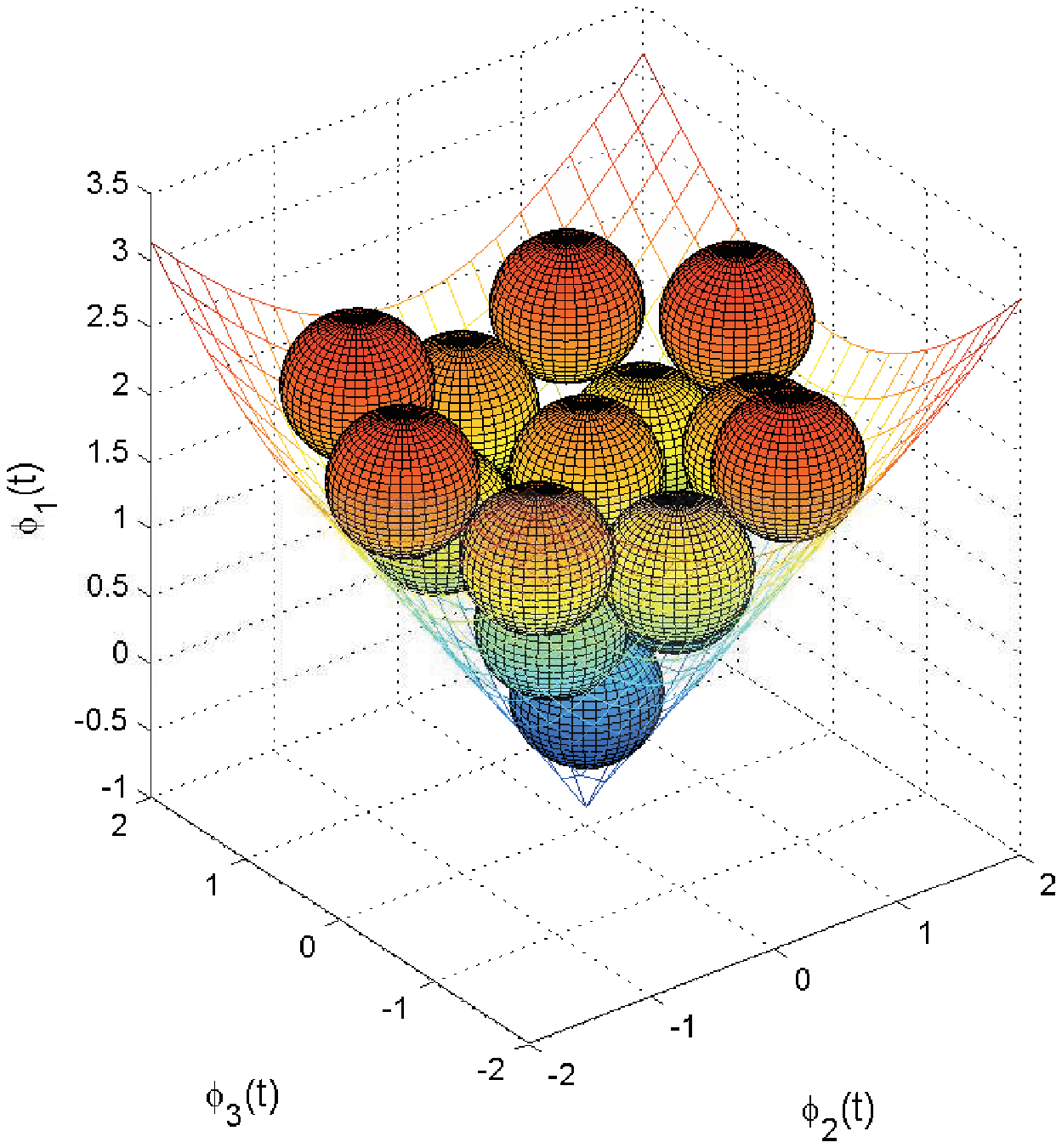} &
\includegraphics[width=0.23\textwidth]{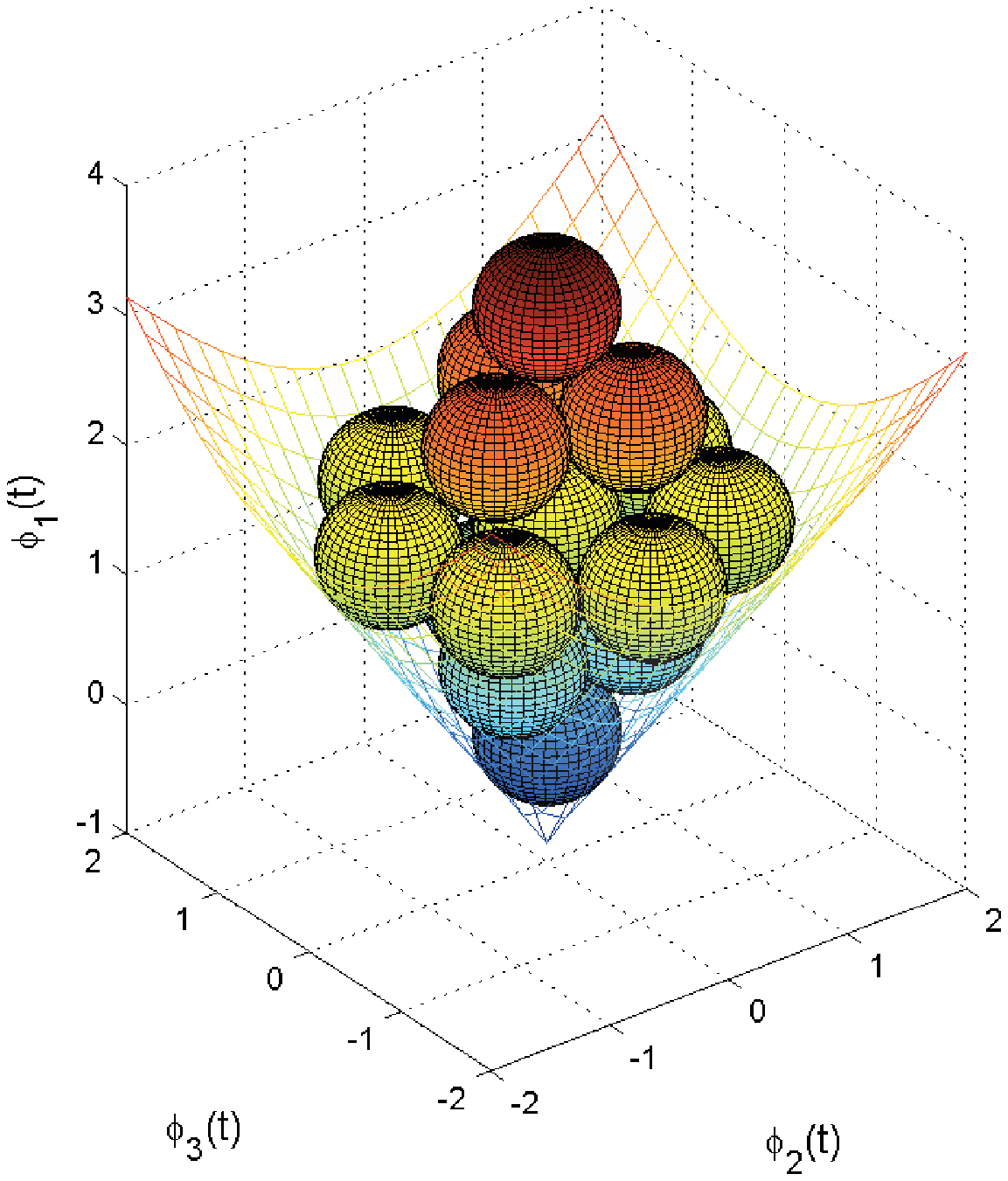} &
\includegraphics[width=0.23\textwidth]{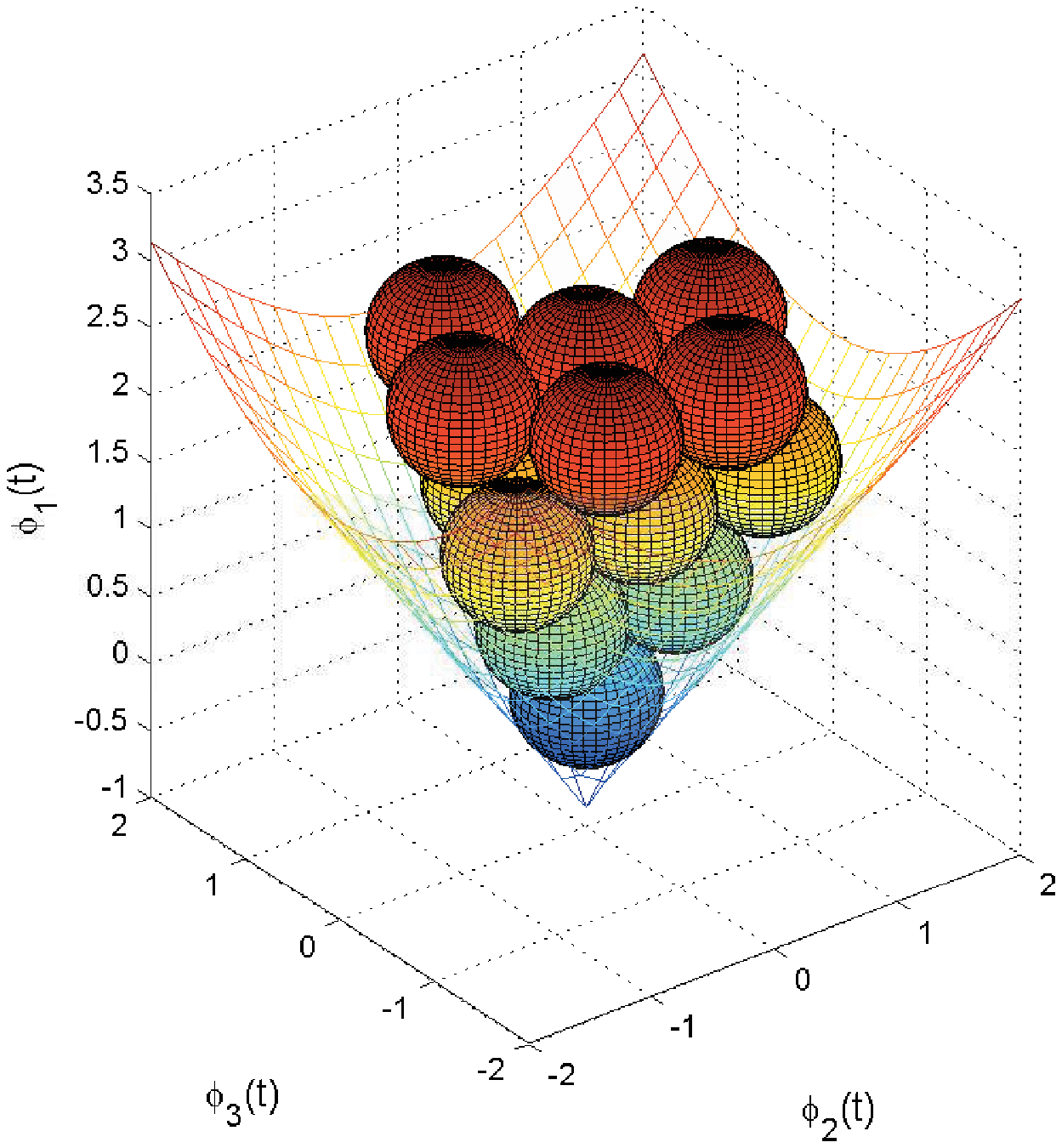}
\end{tabular}
\caption{(left to right): $\mathscr{C}_{\bar P_e,16}$,
 $\mathscr{C}_{\bar P_o,16}$,
  $\mathscr{C}_{\hat P_o,16}$,
  $\mathscr{L}_{16}$.}
\label{16pointscone}

\end{figure*}

\subsection{Optimized Constellations}\label{subsec:optimconst}
In Figs.~\ref{4pointscone}--\ref{16pointscone}, the results of the numerical optimizations are illustrated for $M=4$, $8$, and $16$, for unconstrained sphere packings ($\mathscr{C}$) and lattice codes ($\mathscr{L}$), and for the three power measures. Their coordinates are included in App.~\ref{bestconstapp}. It should be noted that rotations of all constellations about $\phi_1(t)$ do not change the power requirements. As we shall see in Sec~\ref{sec:perf}, the obtained constellation outperform previously known formats.
\begin{conjecture}
All constellations in App.~\ref{bestconstapp} are optimal solutions of~(\ref{minline})--(\ref{cons1}).
\end{conjecture}


\subsubsection{$4$-level Constellations}
The same 4-level constellation provides the lowest $\bar P_e$, $\bar P_o$, and $\hat P_o$ while satisfying the optimization constraints. The geometry of this constellation is a regular tetrahedron where all the spheres, or the constellation points lying at the vertices of this regular tetrahedron, are equidistant from each other.
This constellation is also the result of the optimization constrained to $\Omega \subset \Upsilon \cap A_3$, where the apex of the cone coincides with a point in the $A_3$ lattice and the lattice is oriented such that two lattice basis vectors lie in the plane spanned by $\phi_2(t)$ and $\phi_3(t)$. Since the obtained constellation is optimized for $\bar P_e$, $\bar P_o$, and $\hat P_o$, we will refer to it as $\mathscr{C}_4$ or $\mathscr{L}_4$.

It is a remarkable fact that the vertex angle of the tetrahedron, defined as the apex angle of the circumscribed cone, is exactly $\cos^{-1}(1/3)$, which is equal to the apex angle of the admissible region $\Upsilon$. 
Thus, $\mathscr{C}_4$ fits $\Upsilon$ snugly, in the sense that all constellation points are equidistant from each other and lie on the boundary of $\Upsilon$.
For constellation points regarded as unit-diameter spheres, $\mathscr{C}_4$ can be illustrated as four spheres touching each other and the boundary of a larger cone as shown in Fig.~\ref{4pointscone}. 
This, as we shall see in the next section, makes the modulation format very power-efficient.
This modulation format consists of a zero-level signal and a biased ternary PSK constellation~\cite{Pierce1980,Ekanayake1982}.
In prior work~\cite{Karout2010}, the $\mathscr{C}_4$ format was introduced where it was called on-off phase-shift keying (OOPSK), and in~\cite{Szczerba2011}, it was demonstrated experimentally.
%
%
Other hybrids between amplitude-shift keying and PSK have been studied in~\cite{Essiambre2010} and \cite{Gursoy2009}; however, such modulation formats do not satisfy the nonnegativity constraint of IM/DD channels.

\subsubsection{$8$-level Constellations}
%
The highly symmetric and compact constellation $\mathscr{C}_{\bar P_e,8}$ consists of four central spheres arranged in a tetrahedron and four additional spheres, each touching three spheres in the central tetrahedron. Surprisingly, seven of the eight spheres touch the conical boundary of $\Upsilon$. This modulation format is a hybrid between 2-PAM and two ternary PSK constellations, which are DC-biased differently. The constellation $\mathscr{C}_{\bar P_o,8}$ is the same as $\mathscr{C}_{\bar P_e,8}$ but with the top central sphere moved to the boundary of the admissible region.
The constellation optimized for peak optical power, $\mathscr{C}_{\hat P_o,8}$, consists of two tetrahedra lying on top of each other, where one is reflected and rotated $\pi/3$ about $\phi_1(t)$.

On the other hand, when confining the set of points to a lattice structure, the resulting constellations which provide the lowest $\bar P_e$ and $\bar P_o$ are the same, $\mathscr{L}_{\bar P_e,8}=\mathscr{L}_{\bar P_o,8}$.
However, the lattice-based constellation which is optimized for $\hat P_o$, $\mathscr{L}_{\hat P_o,8}$, is the same as $\mathscr{C}_{\hat P_o,8}$.


\subsubsection{$16$-level Constellations}

The constellations $\mathscr{C}_{\bar P_e,16}$, $\mathscr{C}_{\bar P_o,16}$, and $\mathscr{C}_{\hat P_o,16}$ are not lattice codes; however, the shape obtained could be well justified by the contour plots shown in Fig.~\ref{contour}.
The $\mathscr{C}_{\bar P_o,16}$ constellation contains the constellations $\mathscr{C}_{\bar P_o,8}$ and $\mathscr{C}_{\bar P_e,8}$, whereas $\mathscr{C}_{\hat P_o,16}$ consists of two tetrahedra of which one is reflected and eight spheres lying in between, almost at the same level. The presence of two tetrahedra in the constellations optimized for peak optical power, whether 8- or 16-levels, is due to the fact that these constellations are bounded from above by another cone with the same apex angle as that of $\Upsilon$. This can be inferred from the expression of $\hat P_o$ in (\ref{peakpower})~\cite[Th.~2]{Hranilovic2003}.

However, when only lattice-based structures are considered, the constellations providing the lowest  $\bar P_e$,$~\bar P_o$,$~\hat P_o$ are the same, i.e., $\mathscr{L}_{16}=\mathscr{L}_{\bar P_e,16}=\mathscr{L}_{\bar P_o,16}=\mathscr{L}_{\hat P_o,16}$. From Figs.~\ref{8pointscone}--\ref{16pointscone}, it can be noticed that $\mathscr{L}_{16}$ contains both $\mathscr{C}_{\hat P_o,8}$
and $\mathscr{L}_{\bar P_e,8}$, and 
that the $\mathscr{C}_{4}$ constellation is included in all the obtained constellations.

\begin{conjecture}
For single-subcarrier IM/DD systems, the $\mathscr{C}_{4}$ constellation is included in all optimal constellations with $M \geq 4$.
\end{conjecture}


\subsection{Previously Known Constellations}

Our investigation encompasses some previously best known formats, which are presented after being normalized to unit $d_{\text{min}}$. 
Readers are referred to~\cite{Karout2011LicThesis} for the performance of classical formats such as $M$-PSK and $M$-QAM over IM/DD channels.
At spectral efficiency $\eta=1$ bit/s/Hz (where $R=\log_2 M$), OOK is defined as $\{(0),(1)\}$ in terms of $\phi_1(t)$.

At spectral efficiency $\eta=1.5$ bit/s/Hz, 
 %
a star-shaped 8-QAM~\cite{Essiambre2010} denoted as $\breve 8$-QAM, in which the DC bias is allowed to vary from symbol to symbol, is defined as
$\{ ( 1,\pm 1/2 ,\pm 1/2  ), \allowbreak
((1+\sqrt{3})/\sqrt{2},0,\pm (1+\sqrt{3})/2),\allowbreak
((1+\sqrt{3})/\sqrt{2},\pm (1+\sqrt{3})/2,0) \}$.
At spectral efficiency $\eta=2$ bit/s/Hz, 
nonnegative $4$-PAM is defined as $\{(0),(1),(2),(3)\}$ in terms of $\phi_1(t)$, and 
%
 a version of $16$-QAM denoted as $\breve {16}$-QAM where the DC bias varies from symbol to symbol, is defined as
$\{
(1, \pm 1/2, \pm 1/2),\allowbreak
(\sqrt{5} ,\pm 1/2, \pm 3/2),\allowbreak
(\sqrt{5}, \pm 3/2, \pm 1/2) ,\allowbreak
(3, \pm 3/2, \pm 3/2)
\}$.

\section{Performance Analysis}\label{sec:perf} \vspace{0.2cm}
In this section, we assess the performance of all the modulation formats considered in this work in terms of their symbol error rate (SER) performance, asymptotic power gain versus OOK, which was used as a benchmark in~\cite{Barry1994,Kahn1997}, and spectral efficiency. Sections~\ref{unc:1}--\ref{unc:2} consider uncoded transmission ($R=\log_2 M$) and Sections~\ref{co:1}--\ref{co:2} consider coded transmission ($R=  I(\mathbf{x};\mathbf{y})$).

\subsection{Symbol Error Rate}\label{unc:1}

\begin{figure}
\centering

\footnotesize
\psfrag{aaaaaaaa1}[c][c][0.75][0]{$\mathrm{OOK}$}
 \psfrag{aaaaaaaa2}[c][c][0.75][0]{QPSK }
\psfrag{aaaaaaaa3}[c][c][0.9][0]{$\mathscr{C}_4$ }
\psfrag{aaaaaaaa4}[c][c][0.75][0]{8-QAM}
\psfrag{aaaaaaaa5}[c][c][0.9][0]{$\mathscr{L}_{\bar P_e,8}$}
\psfrag{aaaaaaaa6}[c][c][0.9][0]{$\mathscr{C}_{\bar P_e,8}$}
\psfrag{aaaaaaaa7}[c][c][0.9][0]{$\mathscr{C}_{\bar P_o,8}$}
\psfrag{aaaaaaaa8}[c][c][0.9][0]{$\mathscr{C}_{\hat P_o,8}$}
\psfrag{aaaaaaaa9}[c][c][0.75][0]{4-PAM}
\psfrag{aaaaaaaa10}[c][c][0.75][0]{16-QAM}
\psfrag{aaaaaaaa11}[c][c][0.75][0]{$\mathscr{L}_{16}$}
\psfrag{aaaaaaaa12}[c][c][0.9][0]{$\mathscr{C}_{\bar P_e,16}$}
\psfrag{aaaaaaaa13}[c][c][0.9][0]{$\mathscr{C}_{\bar P_o,16}$}
\psfrag{aaaaaaaa14}[c][c][0.9][0]{$\mathscr{C}_{\hat P_o,16}$}
\psfrag{xxxxxxxx4}[c][c][0.75][0]{$\breve 8$-QAM}
\psfrag{xxxxxxxx10}[c][c][0.75][0]{$\breve {16}$-QAM}
\psfrag{xxxxxxxx5}[c][c][0.75][0]{8-PSK}
\psfrag{aaaaaaaa18}[c][c][0.75][0]{16-PSK}

\psfrag{SER}[c][c][1][0]{$P_s$}
\psfrag{SNREbN0dB}[t][c][1][0]{$\gamma_{ E_b}~\text{[dB]}$}
\psfrag{TbavePosquaredNodB}[t][c][1][0]{$\gamma_{\bar P_o}~\text{[dB]}$}
\psfrag{TbpeakPosquaredNodB}[t][c][1][0]{$\gamma_{\hat P_o}~\text{[dB]}$}
\begin{tabular}{c}
\vspace{1mm}
\includegraphics[scale=0.51]{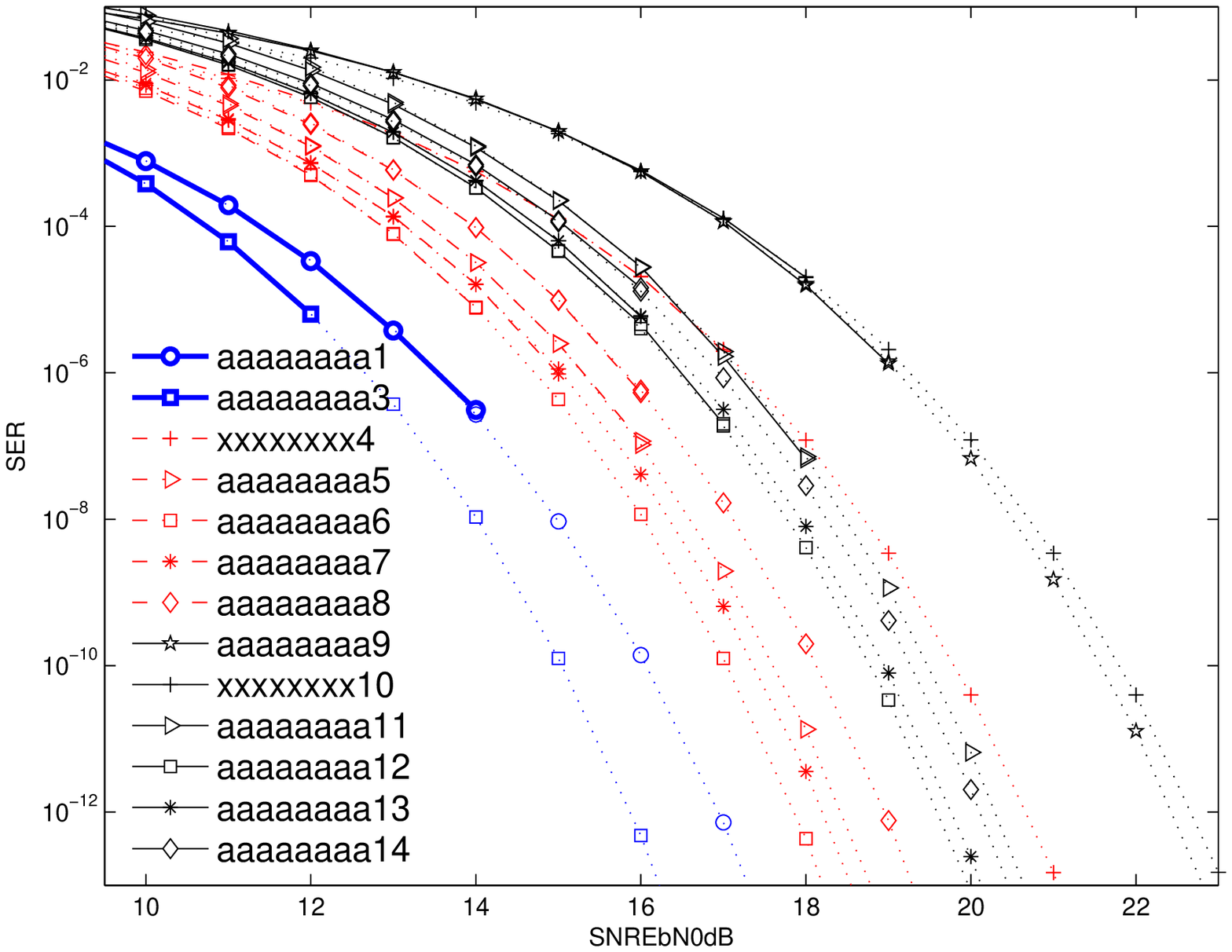}    \\ \vspace{1mm}
\includegraphics[scale=0.51]{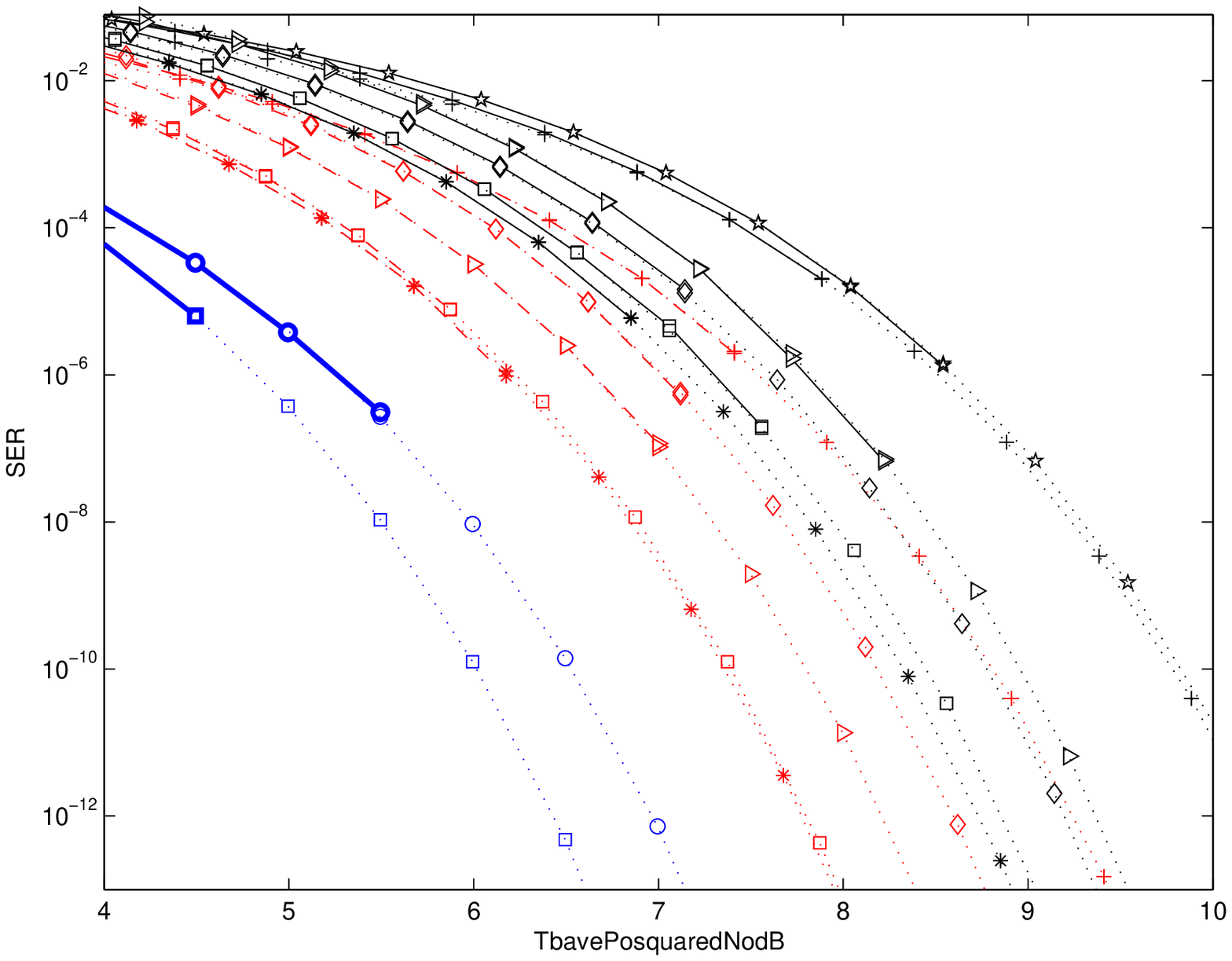} \\
\includegraphics[scale=0.51]{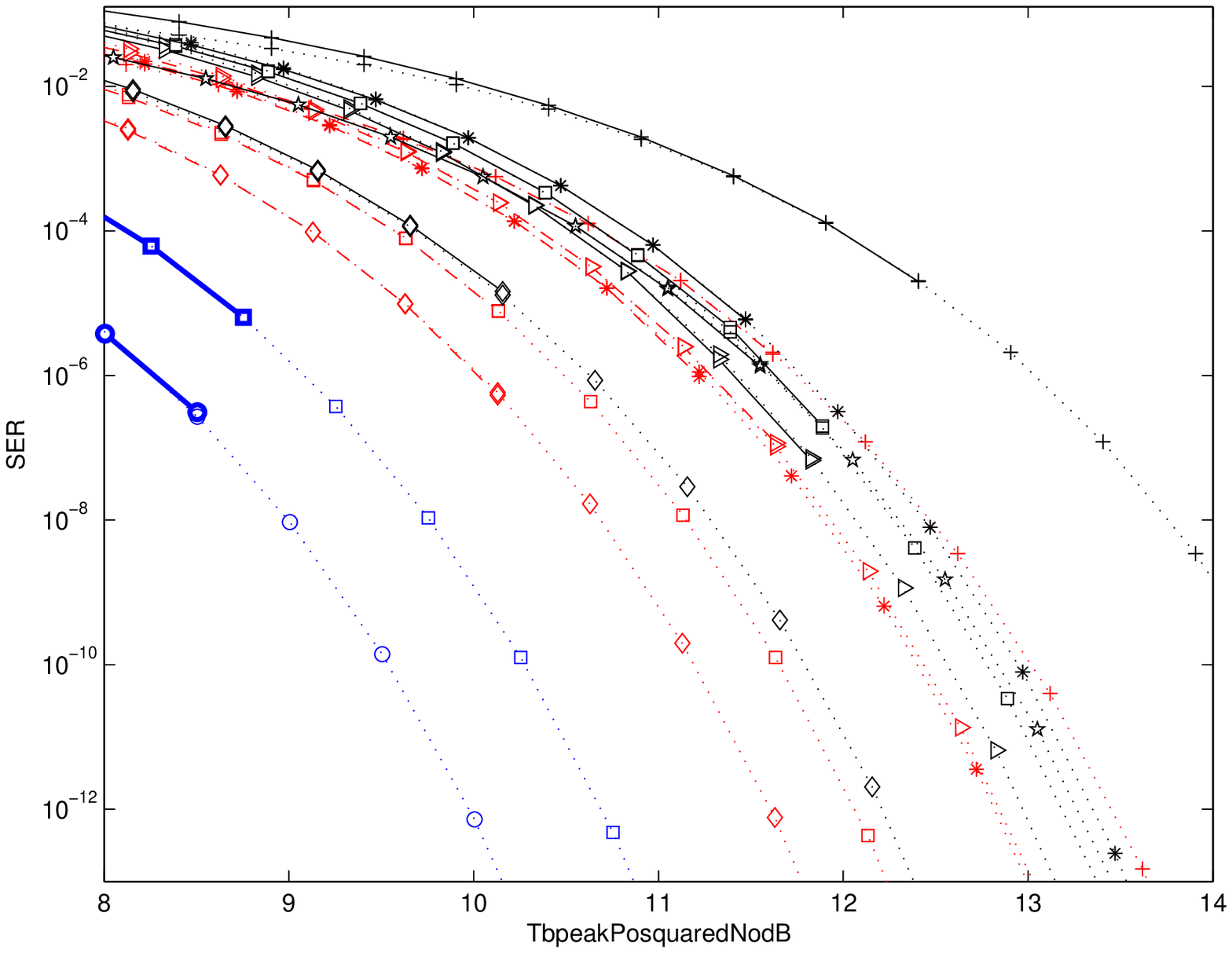} \\
\end{tabular}
\caption{Simulated (solid) and theoretical (dotted) SER for the modulation formats  vs. $\gamma_{E_b}$ (top), $\gamma_{\bar P_o}$ (middle), and $\gamma_{\hat P_o}$ (bottom) without coding.}
\label{serebno}
\end{figure}

For the $4$-level modulation $\mathscr{C}_4$ presented in Fig.~\ref{4pointscone}, deriving the exact theoretical SER is not straight-forward, due to the irregularity of the Voronoi regions. However, it has the same structure as the simplex signal set in~\cite[Sec. 4.1]{Simon1995}, although it is DC-biased to be used in IM/DD systems.
The exact SER of an $M$-ary simplex signal set is~\cite[Eq.~(4.116)]{Simon1995}
 \begin{equation}\label{sersimplex}\begin{split}
   P_s =
   1-   \int_{-\infty}^{\infty} \!
   \Bigg[
   1- ~\mathrm{Q} \Bigg( u \,+ \,&\sqrt{\frac{ 2   E_{s,\text{simplex}}}{ N_0}\frac{ M}{ M-1}}  \Bigg)
   \Bigg]^{M-1}
      \\ &   \cdot   \frac{e^{-u^2/2}}{\sqrt{2\pi}}      \, \mathrm{d}u,
 \end{split}\end{equation}
where $ E_{s,\text{simplex}}$ is the average symbol energy of the zero-mean simplex constellation and 
$\mathrm{Q}(x)=     {1}/{\sqrt{2 \pi}}          \int_{x}^{\infty} \! \exp(- {u^2}/{ 2}) \, \mathrm{d}u$
is the Gaussian $\mathrm{Q}$-function.
For $M=4$, the energy of the zero-mean simplex constellation is $ E_{s,\text{simplex}}=  E_{s}/2=E_b$, where
$E_s$ and $E_b$ are the average symbol and bit energies, respectively, of the $\mathscr{C}_4$ constellation. 
Hence, the exact SER of $\mathscr{C}_4$ is
 \begin{equation}
   P_s=1-
   \int_{-\infty}^{\infty} \!
   \Bigg[
  1-  ~\mathrm{Q} \Bigg( u + \sqrt{ \frac{8}{3} \frac{  E_{b}}{ N_0}}  \Bigg)
    \Bigg]^3
        \frac{e^{-u^2/2}}{\sqrt{2\pi}}      \, \mathrm{d}u.
\end{equation}


For higher-level modulation formats, the standard union bound found in~\cite[Eq.~(4.81)]{Simon1995} is used to approximate the theoretical SER. This union bound can be approximated as
\begin{equation}\label{bound}
    P_{  s   } \approx \frac{2 K }{M }~\mathrm{Q}\left(   \sqrt{\frac{ d_{\text{min}}^2}{2 N_0}}  \right),
\end{equation}
where $K$ is the number of distinct signal pairs $(s_i(t),s_j(t))$ with $i<j$ for which $\int (s_i(t)-s_j(t))^2  \, \mathrm{d}t = d_{\text{min}}^2$. This approximation is tight at high SNR.

Fig.~\ref{serebno} (top) shows the simulated and theoretical SER of the studied modulation formats vs. electrical SNR defined as
\begin{equation}\label{snreb}
    \gamma_{ E_b} =10 \log_{10}\frac{ E_b}{N_0} ~~\textrm{[dB]}.
\end{equation}
As expected, all the modulation formats which are optimized for $\bar P_e$ outperform the other formats at the same spectral efficiency.
For spectral efficiency $\eta=1$ bit/s/Hz (blue),
$\mathscr{C}_4$ has a
0.86 dB average electrical power gain over OOK
to achieve $P_s= 10^{-6}$.
For $\eta=1.5$ bit/s/Hz (red),
$\mathscr{C}_{\bar P_e,8}$ has a
0.31 dB gain over $\mathscr{C}_{\bar P_o,8}$,
0.58 dB gain over $\mathscr{L}_{\bar P_e,8}$, 
1.09 dB gain over $\mathscr{C}_{\hat P_o,8}$, 
and 2.55 dB gain over $\breve 8$-QAM
to achieve $P_s= 10^{-6}$.
For $\eta=2$ bit/s/Hz (black),
$\mathscr{C}_{\bar P_e,16}$ has a
0.15 dB gain over $\mathscr{C}_{\bar P_o,16}$,
0.47 dB gain over $\mathscr{C}_{\hat P_o,16}$,
0.74 dB gain over $\mathscr{L}_{16}$,
2.65 dB gain over 4-PAM,
and 2.80 dB gain over $\breve {16}$-QAM.
%
%
The modulation formats optimized for $\bar P_e$ and $\bar P_o$ are close in performance to those optimized for $\hat P_o$ and to the lattice-based formats.


In order to compare modulation formats in terms of their average optical power requirements, we define the average optical SNR as
\begin{equation}\label{snravePo}
    \gamma_{\bar P_o} = 10 \log_{10}\frac{  \bar P_o}{c\sqrt{R_b N_0}}~~\textrm{[dB]}
\end{equation}
in a similar fashion as in~\cite[Eq.~(5)]{Kahn1997}. 
Using~(\ref{aveelesig}),~(\ref{opticalpowersig}),~(\ref{snreb}), $rc=1$, and $E_s=E_b R_b T_s$, this expression can be written as
\begin{equation}\label{exp:gamma:po}
 \gamma_{\bar P_o} = \frac{1}{2} \gamma_{ E_b} + 10 \log_{10} \frac{\mathbb{E} [s_{I,1}]}{\sqrt{\mathbb{E}[ \|\mathbf{s}_I \|^2]}},
\end{equation}
where the first term depends on the regular (electrical) SNR and the second depends only on the constellation geometry.
Fig.~\ref{serebno} (middle) shows the SER plotted vs. $\gamma_{\bar P_o}$. Quite obviously, the modulation formats optimized for $\bar P_o$ perform better than the rest.
For $\eta=1$ bit/s/Hz, $\mathscr{C}_4$ has a
0.43 dB average optical power gain over OOK
to achieve an SER of $10^{-6}$.
For $\eta=1.5$ bit/s/Hz, $\mathscr{C}_{\bar P_o,8}$ has a
0.04 dB gain over $\mathscr{C}_{\bar P_e,8}$,
0.46 dB gain over $\mathscr{L}_{\bar P_e,8}$, 
0.84 dB gain over $\mathscr{C}_{\hat P_o,8}$, 
and 1.35 dB gain over $\breve 8$-QAM
to achieve $P_s= 10^{-6}$.
For $\eta=2$ bit/s/Hz, $\mathscr{C}_{\bar P_o,16}$ has a
0.13 dB gain over $\mathscr{C}_{\bar P_e,16}$,
0.45 dB gain over $\mathscr{C}_{\hat P_o,16}$,
0.66 dB gain over $\mathscr{L}_{16}$,
1.36 dB gain over $\breve {16}$-QAM,
and 1.44 dB gain over 4-PAM
to achieve $P_s= 10^{-6}$.
It can be noticed that the 8-level modulation formats optimized for average electrical and optical power are very close in performance. 

In a similar fashion as $\gamma_{\bar P_o}$, we define the peak optical SNR
\begin{equation}\label{snravePo}
    \gamma_{\hat P_o} = 10 \log_{10}\frac{  \hat P_o}{ c \sqrt{R_b N_0}} ~~\textrm{[dB]}
\end{equation}
in order to assess the different modulation formats under study in terms of their peak optical power requirements. 
Using~(\ref{aveelesig}),~(\ref{peakpower}),~(\ref{snreb}), $rc=1$, and $E_s=E_b R_b T_s$, this expression can be written as
\begin{equation}\label{exp:gamma::hat:po}
 \gamma_{\hat P_o} = \frac{1}{2} \gamma_{ E_b} + 10 \log_{10} \frac{\max_{i} \left\{s_{i,1} +\sqrt{2(s_{i,2}^2+ s_{i,3}^2)} ~\right\}}{\sqrt{\mathbb{E}[ \|\mathbf{s}_I \|^2]}},
\end{equation}
where the first term depends on the regular (electrical) SNR and the second depends only on the constellation geometry.
Fig.~\ref{serebno} (bottom) shows the SER plotted vs. $\gamma_{\hat P_o}$. It is clear that the modulation formats optimized for peak optical power outperform the other formats.
For $\eta=1$ bit/s/Hz, OOK has a
0.82 dB peak optical power gain over $\mathscr{C}_4$
to achieve $P_s= 10^{-6}$.
For $\eta=1.5$ bit/s/Hz, $\mathscr{C}_{\hat P_o,8}$ has a 
0.46 dB gain over $\mathscr{C}_{\bar P_e,8}$,
1.20 dB gain over $\mathscr{C}_{\bar P_o,8}$,
1.25 dB gain over $\mathscr{L}_{\bar P_e,8}$, 
and 1.72 dB gain over $\breve 8$-QAM
to achieve an SER of $10^{-6}$.
For $\eta=2$ bit/s/Hz, $\mathscr{C}_{\hat P_o,16}$ has a
0.81 dB peak optical power gain over $\mathscr{L}_{16}$,
0.98 dB gain over 4-PAM,
1.00 dB gain over $\mathscr{C}_{\bar P_e,16}$,
1.16 dB gain over $\mathscr{C}_{\bar P_o,16}$,
and 2.42 dB gain over $\breve {16}$-QAM
to achieve $P_s= 10^{-6}$.
Overall, modulation formats optimized for $\hat P_o$ perform well in average-power limited systems. In~\cite{Karlsson2011}, a similar conclusion was reached for the case of coherent optical systems.


\begin{figure}
\centering
\footnotesize
\psfrag{aaaaaaaaa1}[c][c][0.75][0]{$\mathrm{OOK}$}
 \psfrag{aaaaaaaaa2}[c][c][0.75][0]{QPSK }
\psfrag{aaaaaaaaa3}[c][c][0.9][0]{$\mathscr{C}_4$ }
\psfrag{aaaaaaaaa4}[c][c][0.75][0]{8-QAM}
\psfrag{aaaaaaaaa5}[c][c][0.9][0]{$\mathscr{L}_{\bar P_e,8}$}
\psfrag{aaaaaaaaa6}[c][c][0.9][0]{$\mathscr{C}_{\bar P_e,8}$}
\psfrag{aaaaaaaaa7}[c][c][0.9][0]{$\mathscr{C}_{\bar P_o,8}$}
\psfrag{aaaaaaaaa8}[c][c][0.9][0]{$\mathscr{C}_{\hat P_o,8}$} 
\psfrag{aaaaaaaaa9}[c][c][0.75][0]{4-PAM}
\psfrag{aaaaaaaaa10}[c][c][0.75][0]{16-QAM}
\psfrag{aaaaaaaaa11}[c][c][0.75][0]{$\mathscr{L}_{16}$}
\psfrag{aaaaaaaaa12}[c][c][0.9][0]{$\mathscr{C}_{\bar P_e,16}$}
\psfrag{aaaaaaaaa13}[c][c][0.9][0]{$\mathscr{C}_{\bar P_o,16}$}
\psfrag{aaaaaaaaa14}[c][c][0.9][0]{$\mathscr{C}_{\hat P_o,16}$}
\psfrag{raxis}[t][c][1][0]{$\eta$ [bit/s/Hz]}
\psfrag{avepegain}[c][c][1][0]{$\bar{P}_{e,{gain}}~\text{[dB]} $}
\psfrag{avepogain}[c][c][1][0]{$\bar{P}_{o,{gain}}~\text{[dB]}$}
\psfrag{peakpogain}[c][c][1][0]{$\hat{P}_{o,{gain}}~\text{[dB]}$}

\psfrag{xxxxxxxxx4}[c][c][0.75][0]{$\breve 8$-QAM}
\psfrag{xxxxxxxxx10}[c][c][0.75][0]{$\breve {16}$-QAM}
\psfrag{xxxxxxxxx5}[c][c][0.75][0]{8-PSK}
\psfrag{aaaaaaaa18}[c][c][0.75][0]{16-PSK}

\begin{tabular}{c}
\includegraphics[scale=0.51]{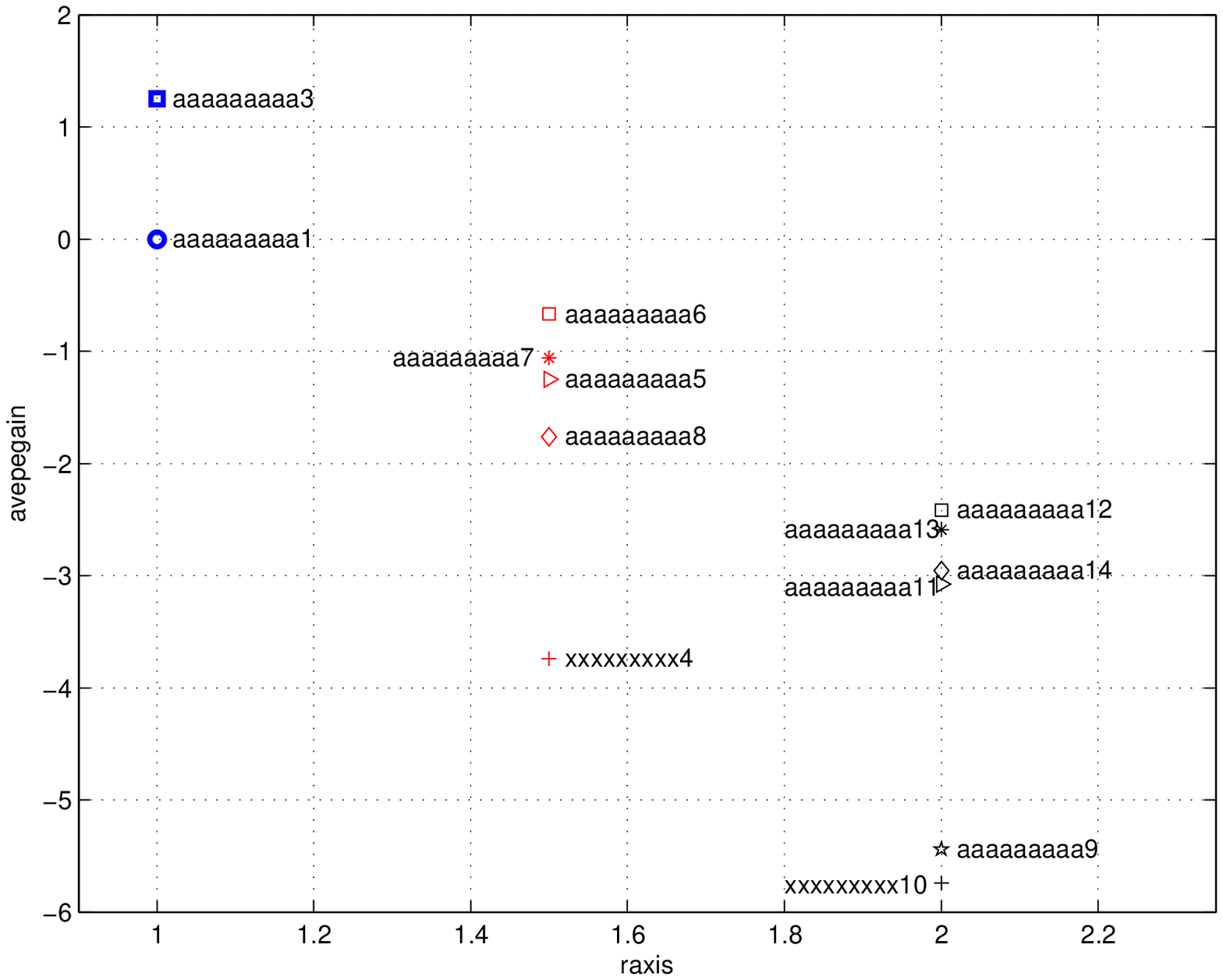}  \\\\
\includegraphics[scale=0.51]{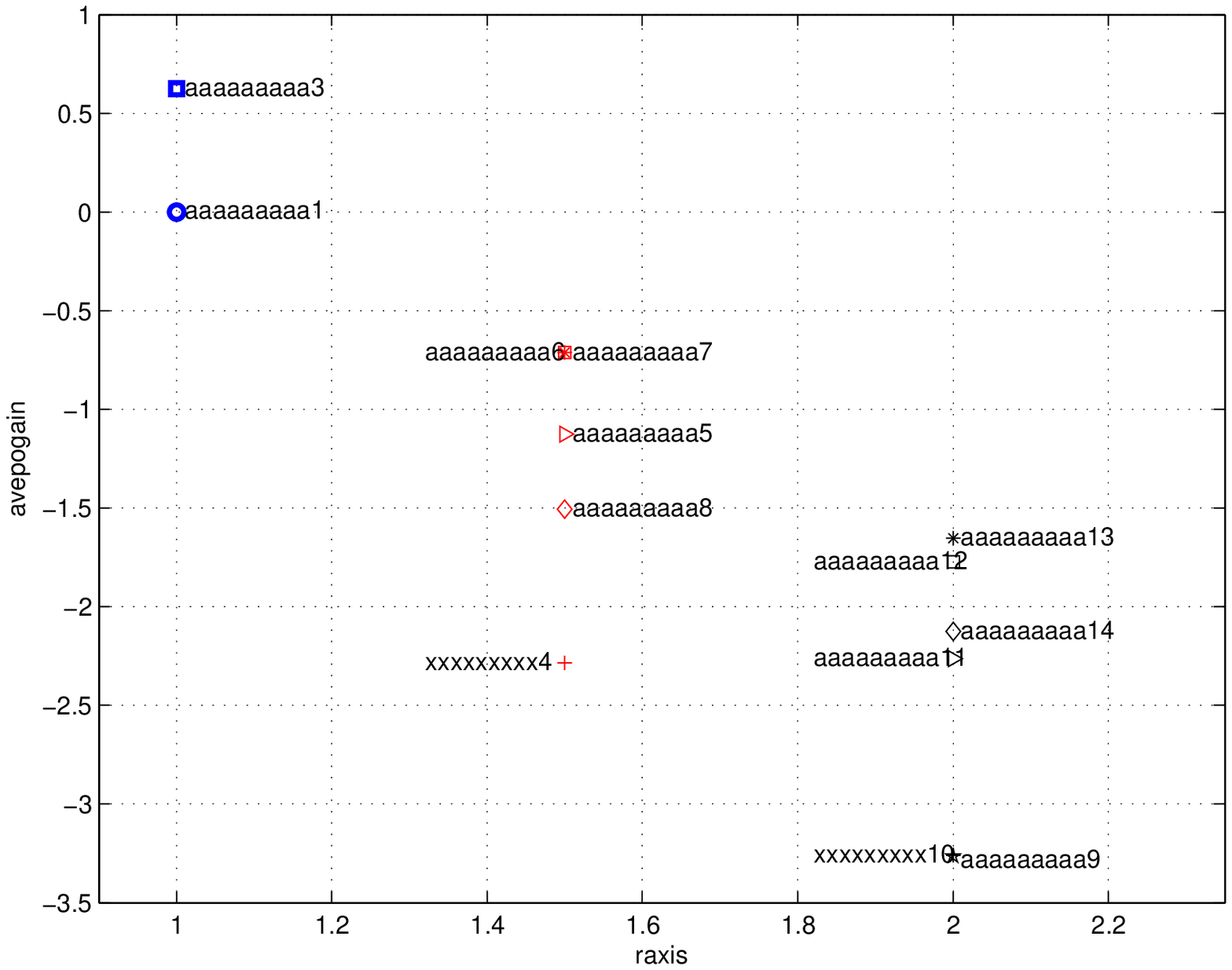} \\\\
\includegraphics[scale=0.51]{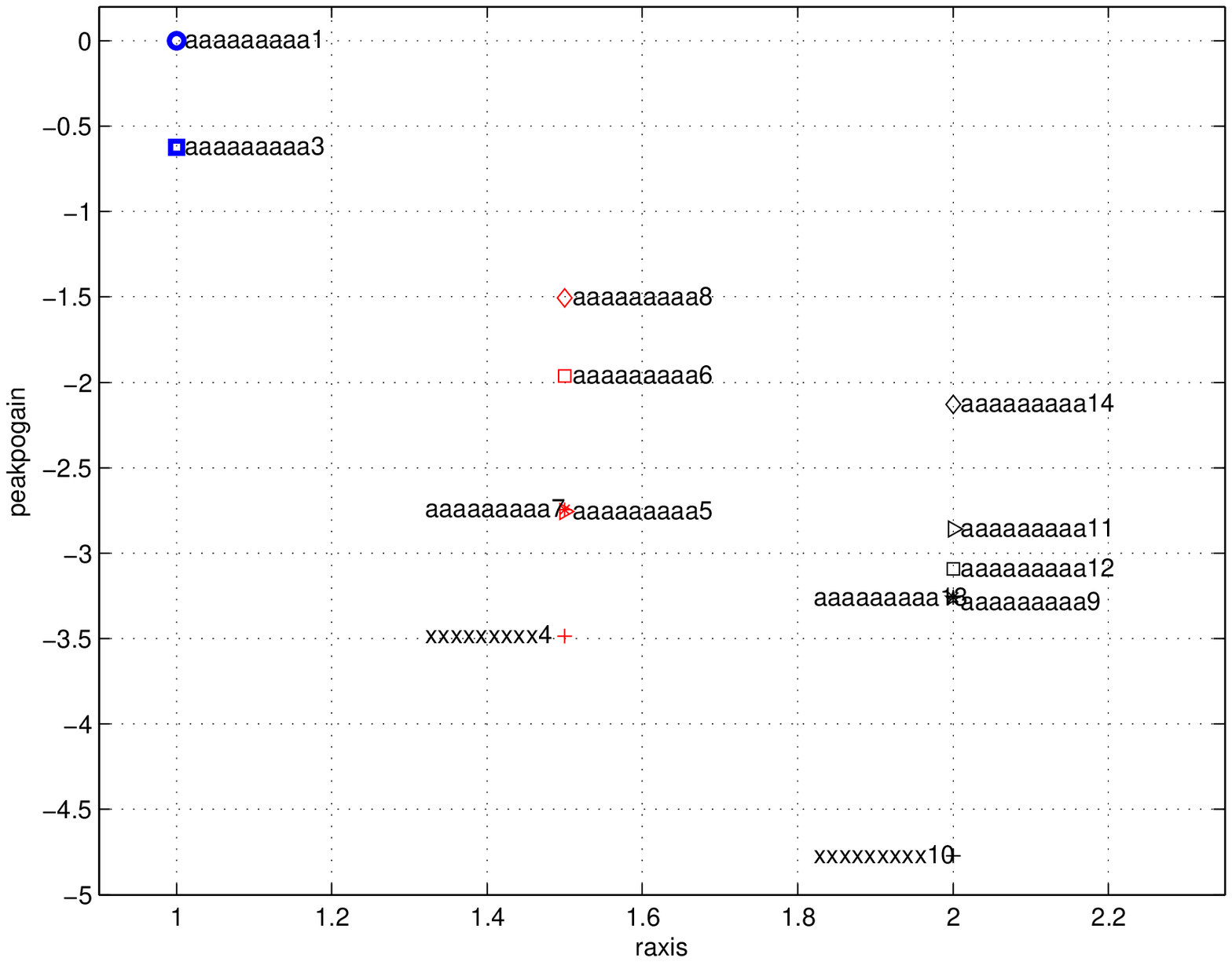} \\ \\
\end{tabular}
\caption{Average electrical (top), average optical (middle), and peak optical (bottom) power gain of the various modulation formats vs. OOK without coding.}
\label{powergainfig}

\end{figure}

\subsection{Asymptotic Power Efficiency}\label{unc:2}

At asymptotically high SNR, the performance difference in dB between the different modulation formats approaches constant values. We use OOK as a benchmark for power-efficiency on IM/DD channels as in~\cite{Barry1994,Kahn1997}.

Fig.~\ref{powergainfig} (top) presents the average electrical power gain
\begin{equation}
    \bar{P}_{e,{gain}}=10 \log_{10} \frac{\bar{P}_{e,{ \text{OOK} }}}{\bar{P}_e} ~~\textrm{[dB]}
\end{equation}
of a modulation format in comparison to OOK in order to achieve the same error rate performance at asymptotically high SNR,
where $\bar{P}_{e,{ \text{OOK} }}$ and ${\bar{P}_e}$ are the average electrical power of OOK and the modulation format under study, respectively. 
For spectral efficiency $\eta=1$ bit/s/Hz, $\mathscr{C}_4$ has a 1.25 dB average electrical power gain over OOK.
The overall trend at asymptotically high SNR is similar to the comparison of average electrical SNR $\gamma_{ E_b}$ in Sec.~\ref{unc:1} to achieve $P_s=10^{-6}$.
%
For $\eta=1.5$ bit/s/Hz, the 8-level modulation format optimized for $\bar P_e$, i.e., $\mathscr{C}_{\bar P_e,8}$, has the smallest average electrical power penalty of $0.67$ dB compared to OOK.
The asymptotic average electrical power gain of $\mathscr{C}_{\bar P_e,8}$ over other formats is larger than the gains at $P_s=10^{-6}$ with the exception of having a similar gain when compared to
$\mathscr{L}_{\bar P_e,8}$ and
 $\mathscr{C}_{\hat P_o,8}$ where both are lattice-based formats.
%
For $\eta=2$ bit/s/Hz, $\mathscr{C}_{\bar P_e,16}$ has a 2.42 dB penalty with respect to OOK at asymptotically high SNR.
Its asymptotic electrical power gain is larger than the gains at $P_s=10^{-6}$ except when compared to $\mathscr{L}_{16}$ where the gain at asymptotically high SNR is smaller by 0.08 dB.

%
%

Fig.~\ref{powergainfig} (middle) shows the average optical power gain
\begin{equation}
    \bar{P}_{o,{gain}}=10 \log_{10} \frac{\bar{P}_{o,{\text{OOK}}}}{\bar{P}_o} ~~\textrm{[dB]}
\end{equation}
of the modulation formats under study with respect to OOK at asymptotically high SNR, plotted versus their spectral efficiencies.
The average optical power of OOK is $\bar{P}_{o,{ \text{OOK} }}$, whereas the average optical power of the modulation format under study is denoted as ${\bar{P}_o}$.
For $\eta=1$ bit/s/Hz, $\mathscr{C}_4$ offers 0.62 dB average optical power gain over OOK.
The asymptotic average optical power gain of $\mathscr{C}_4$ over OOK is larger than the gain at $P_s=10^{-6}$ in Sec.~\ref{unc:1}.
%
%
%
For $\eta=1.5$ bit/s/Hz, $\mathscr{C}_{\bar P_o,8}$ has a 0.71 dB average optical power penalty compared to OOK to achieve the same error rate at asymptotically high SNR. 
Furthermore, the asymptotic average optical power gains of $\mathscr{C}_{\bar P_o,8}$ are larger than the gains when compared with other formats to achieve $P_s=10^{-6}$. Exceptions are when compared to $\mathscr{L}_{\bar P_e,8}$ and $\mathscr{C}_{\hat P_o,8}$, where the gains are smaller. Another observation is that the performance of $\mathscr{C}_{\bar P_o,8}$ becomes similar to that of $\mathscr{C}_{\bar P_e,8}$.
%
%
For $\eta=2$ bit/s/Hz, $\mathscr{C}_{\bar P_o,16}$ has a
1.65 dB penalty with respect to OOK at asymptotically high SNR. 
Compared to the average optical SNR performance gains at $P_s=10^{-6}$, the asymptotic gains are larger except the gain over $\mathscr{C}_{\bar P_e,16}$ and $\mathscr{L}_{16}$.

If peak power is the limiting factor in a communication system, Fig.~\ref{powergainfig} (bottom) shows the performance of the various modulation formats with respect to their peak optical power gain with respect to OOK
\begin{equation}
    \hat{P}_{o,{gain}}=10 \log_{10} \frac{\hat{P}_{o,{\text{OOK}}}}{\hat{P}_o} ~~\textrm{[dB]},
\end{equation}
where $\hat{P}_{o,{ \text{OOK} }}$ and ${\hat{P}_o}$ denote the peak optical power of OOK and the modulation format under study, respectively.
For $\eta=1$ bit/s/Hz,
OOK has the best performance. It has a peak optical power gain of
0.62 dB over $\mathscr{C}_4$. The asymptotic peak optical power gain of OOK compared to $\mathscr{C}_4$ is less than the gain in peak optical SNR to achieve $P_s=10^{-6}$. In other words, the peak optical performance of $\mathscr{C}_4$ gets closer to OOK at asymptotically high SNR.
%
%
%
For $\eta=1.5$ bit/s/Hz, $\mathscr{C}_{\hat P_o,8}$ has the smallest penalty of 
1.51 dB compared to OOK.
The asymptotic peak optical power gain of $\mathscr{C}_{\hat P_o,8}$ over other formats is larger than the gain reported in Sec.~\ref{unc:1}. The exception is the performance of $\mathscr{C}_{\hat P_o,8}$ when compared to $\mathscr{C}_{\bar P_e,8}$ and $\mathscr{L}_{\bar P_e,8}$, where the asymptotic peak optical power gain is similar to that at $P_s=10^{-6}$.
%
%
%
For $\eta=2$ bit/s/Hz, $\mathscr{C}_{\hat P_o,16}$ outperforms the other modulation formats with a penalty of 2.13 dB with respect to OOK at asymptotically high SNR.
Even though the asymptotic peak optical power gain of $\mathscr{C}_{\hat P_o,16}$ over most other formats is larger than that at $P_s=10^{-6}$, it is smaller when $\mathscr{C}_{\hat P_o,16}$ is compared to $\mathscr{L}_{16}$, $\mathscr{C}_{\bar P_e,16}$, and $\mathscr{C}_{\bar P_o,16}$.
So far, all the differences between the gains to achieve $P_s=10^{-6}$ and the gains at asymptotically high SNR did not change the order of which formats have a better performance.
However, 4-PAM at asymptotically high SNR is now worse than $\mathscr{C}_{\bar P_e,16}$ and $\mathscr{C}_{\bar P_o,16}$ in terms of peak optical power.
%
Finally, the trend that can be observed in Fig.~\ref{powergainfig} (bottom) is that the gap in performance between the modulation formats optimized for peak optical power and the rest of the formats under study gets larger with higher spectral efficiencies.

\begin{figure}
\centering
\footnotesize
\psfrag{aaaaaaa1}[c][c][0.75][0]{OOK}
 \psfrag{aaaaaaa2}[c][c][0.75][0]{QPSK }
\psfrag{aaaaaaa3}[c][c][0.9][0]{$\mathscr{C}_4$ }
\psfrag{aaaaaaa4}[c][c][0.75][0]{8-QAM}
 \psfrag{aaaaaaa5}[c][c][0.75][0]{$\mathscr{L}_{\bar P_e,8}$}
\psfrag{aaaaaaa6}[c][c][0.9][0]{$\mathscr{C}_{\bar P_e,8}$}
\psfrag{aaaaaaa7}[c][c][0.9][0]{$\mathscr{C}_{\bar P_o,8}$}
\psfrag{aaaaaaa8}[c][c][0.75][0]{$\mathscr{C}_{\hat P_o,8}$}
\psfrag{aaaaaaa9}[c][c][0.75][0]{4-PAM}
\psfrag{aaaaaaa10}[c][c][0.75][0]{16-QAM}
\psfrag{aaaaaaa11}[c][c][0.75][0]{$\mathscr{L}_{16}$}
\psfrag{aaaaaaa12}[c][c][0.9][0]{$\mathscr{C}_{\bar P_e,16}$}
\psfrag{aaaaaaa13}[c][c][0.9][0]{$\mathscr{C}_{\bar P_o,16}$}
\psfrag{aaaaaaa14}[c][c][0.9][0]{$\mathscr{C}_{\hat P_o,16}$}
\psfrag{xxxxxxx4}[c][c][0.75][0]{$\breve 8$-QAM}
\psfrag{xxxxxxx5}[c][c][0.75][0]{8-PSK}
\psfrag{xxxxxxx10}[c][c][0.75][0]{$\breve {16}$-QAM}
\psfrag{aaaaaaa18}[c][c][0.75][0]{16-PSK}
\psfrag{sssssss101}[c][c][0.75][0]{$\mathscr{E}_4$}
\psfrag{sssssss102}[c][c][0.75][0]{$\mathscr{E}_8$}
\psfrag{sssssss103}[c][c][0.75][0]{$\mathscr{E}_{16}$}
\psfrag{ebnodb}[t][c][1][0]{$\gamma_{ E_b}~\text{[dB]}$}
\psfrag{MI}[c][c][1][0]{$\eta$ [bit/s/Hz]}
\psfrag{TbavePosquaredNodB}[t][c][1][0]{$\gamma_{\bar P_o}~\text{[dB]}$}
\psfrag{TbpeakPosquaredNodB}[t][c][1][0]{$\gamma_{\hat P_o}~\text{[dB]}$}

\begin{tabular}{c}
\includegraphics[scale=0.51]{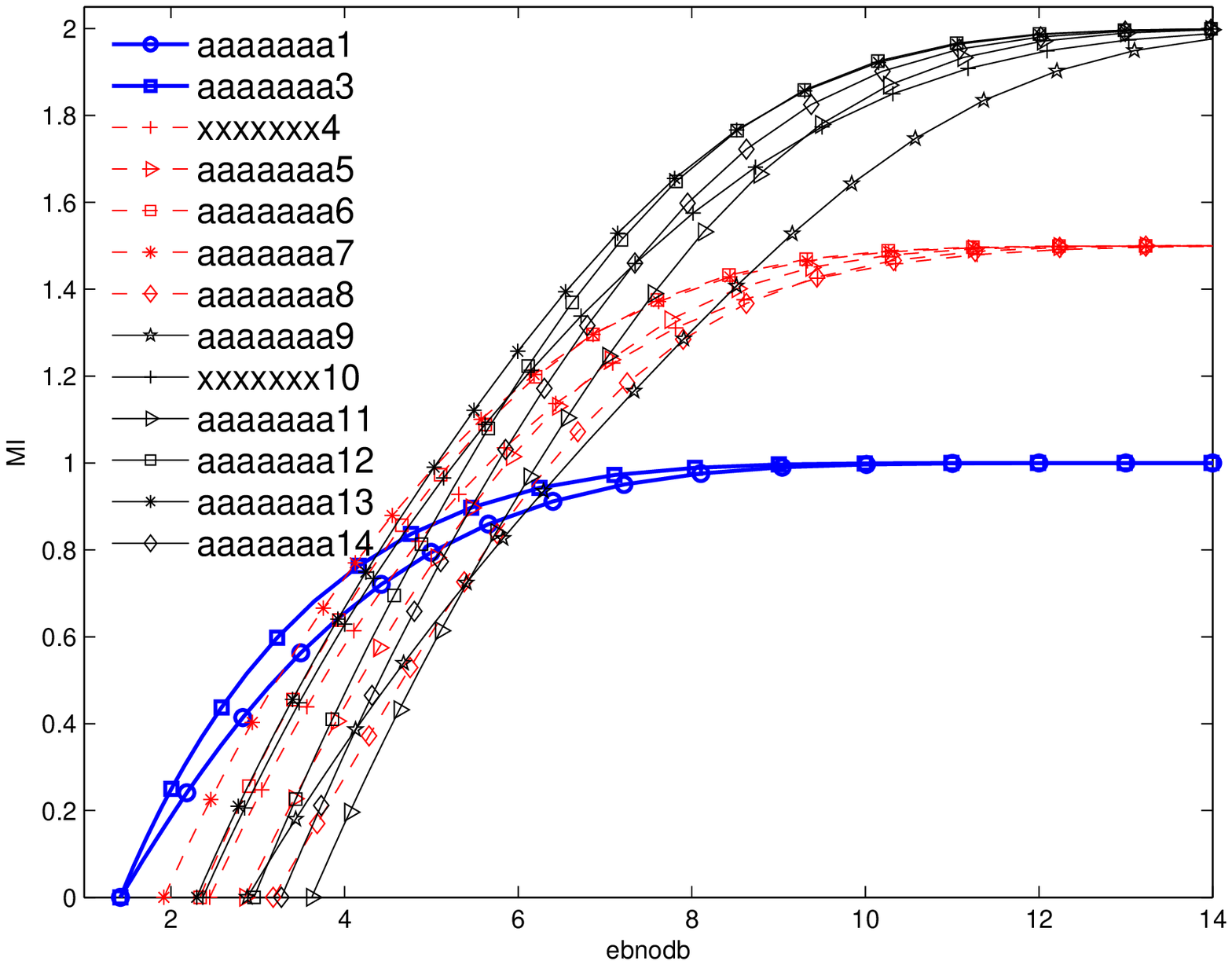} \\ \\
\includegraphics[scale=0.51]{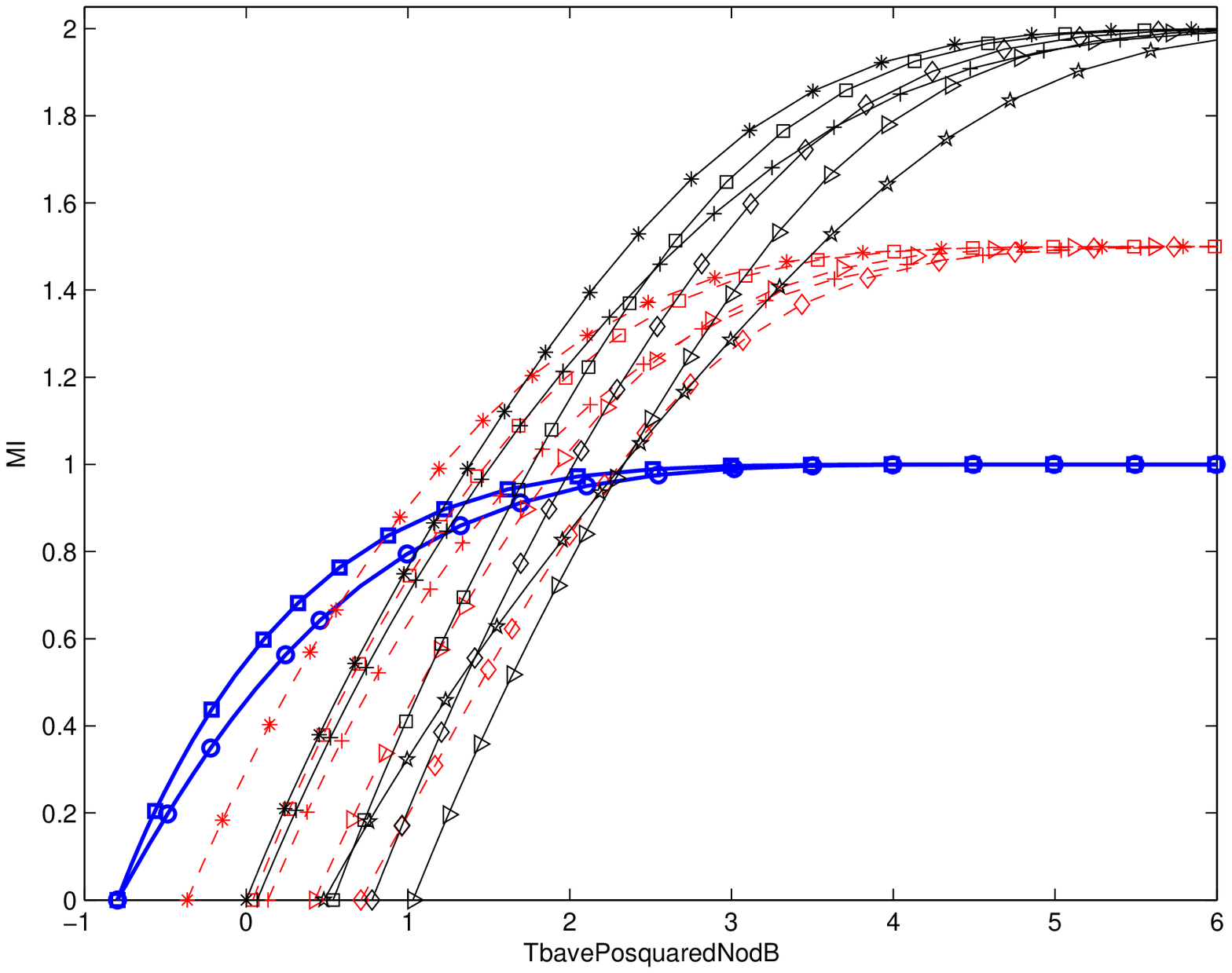}\\ \\
\includegraphics[scale=0.51]{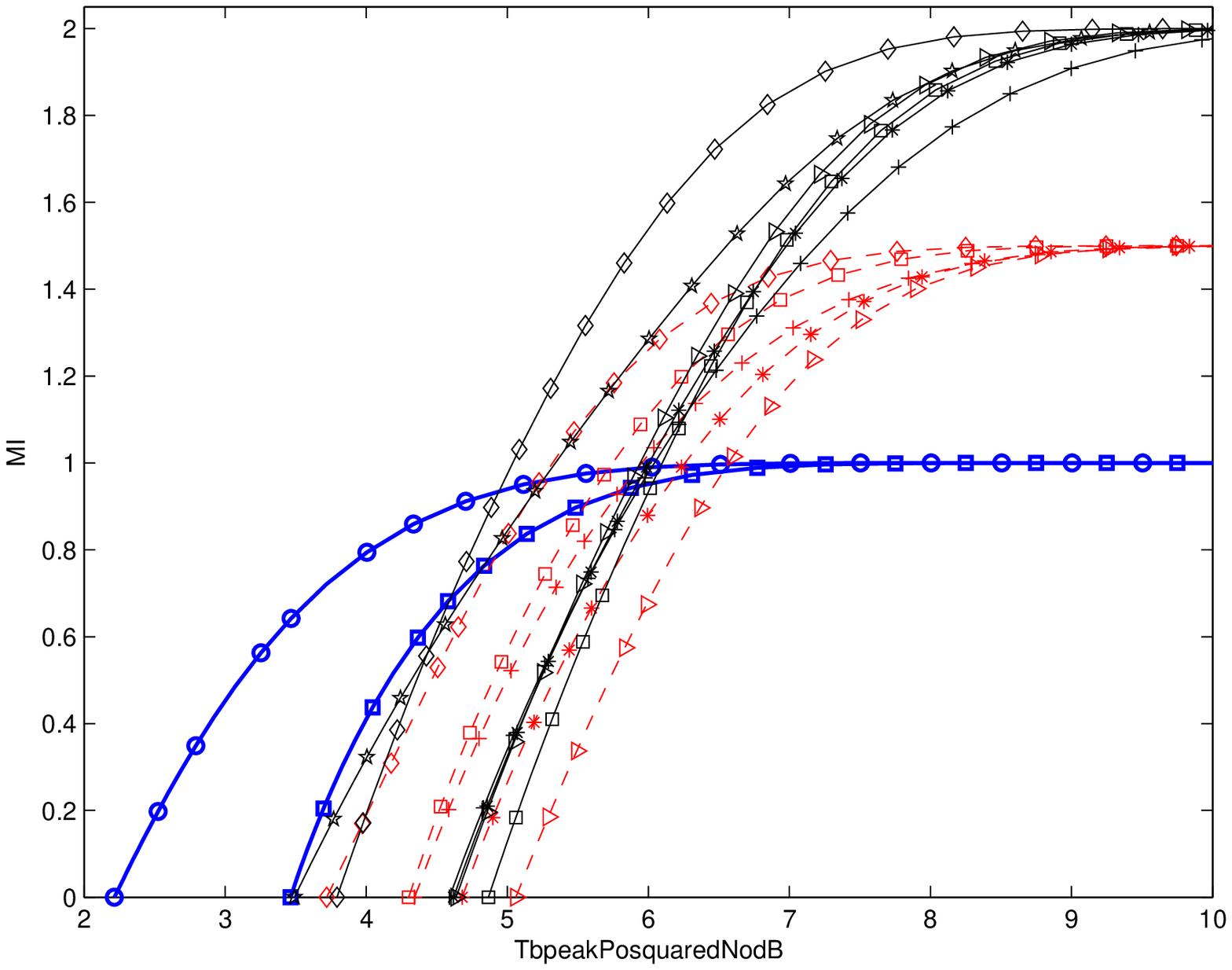}
\end{tabular}
\caption{Spectral efficiency vs. $\gamma_{ E_b}$ (top), $\gamma_{\bar P_o}$ (middle), and $\gamma_{\hat P_o}$ (bottom) in the presence of capacity-achieving codes.}
\label{mi}
\end{figure}

\subsection{Mutual Information vs. SNR}\label{co:1}
Fig.~\ref{mi} depicts the spectral efficiency where $R=  I(\mathbf{x};\mathbf{y})$ of the modulation formats vs. the different SNR measures, i.e., average electrical, average optical, and peak optical SNR. 
We study the same modulation formats as before, designed to minimize the uncoded SER, although they are now evaluated in a coding context.
The mutual information $I(\mathbf{x};\mathbf{y})$ for a fixed constellation and input distribution gives a lower bound on the rate in bits per symbol at which information can be sent with arbitrarily low probability of error, i.e., with the use of optimal coding. Throughout this paper, we consider uniform input distributions.

In Fig.~\ref{mi} (top), the spectral efficiency is plotted vs. $\gamma_{ E_b}$. 
The best format in terms of $\gamma_{ E_b}$ is $\mathscr{C}_4$ for $0 <  \eta< 0.74$, $\mathscr{C}_{\bar P_o,8}$ for $0.74 < \eta <0.99$, and $\mathscr{C}_{\bar P_o,16}$ for $\eta > 0.99$.
It is surprising that the formats optimized for $\bar P_o$ outperform the formats optimized for $\bar P_e$, when compared in terms of $\gamma_{ E_b}$. This is different from the uncoded case in Sec.~\ref{unc:1}--\ref{unc:2}.
 The $\breve 8$-QAM and $\mathscr{L}_{\bar P_e,8}$ formats have a better performance than $\mathscr{C}_{\hat P_o,8}$.
%
Other observations are that $\mathscr{C}_{\hat P_o,16}$ outperforms $\mathscr{L}_{16}$, and 4-PAM 
has the lowest performance.


In a similar fashion as above, Fig.~\ref{mi} (middle) shows the spectral efficiency $\eta$ plotted vs. the SNR $\gamma_{\bar P_o}$ for the same modulation formats.
The best format is $\mathscr{C}_4$ for $ 0< \eta <0.83$, $\mathscr{C}_{\bar P_o,8}$ for $0.83<\eta<1.18$, and $\mathscr{C}_{\bar P_o,16}$ for $\eta >1.18$. 
As opposed to the performance vs. $\gamma_{ E_b}$, the gap between the formats optimized for average optical power and those for average electrical power is larger.
It is interesting that the modulation formats optimized for average optical power perform better than the rest for systems which are limited by either average electrical or optical power.


Fig.~\ref{mi} (bottom) presents the spectral efficiency vs. $\gamma_{\hat P_o}$. But here, the story is a bit different.
The best format is OOK for $ 0< \eta <0.93$, and $\mathscr{C}_{\hat P_o,16}$ above that.
An interesting observation is that
%
4-PAM performs better than $\mathscr{C}_{\hat P_o,16}$ for $0<\eta<0.56$.
Furthermore, modulation formats optimized for peak power perform better in average-power limited systems than when using formats optimized for average power in peak-power limited systems.


\subsection{Mutual Information in the Wideband Regime}\label{co:2}

It is apparent from the results in Sec.~\ref{co:1} that constellations optimized for a minimum-distance criterion (i.e., uncoded transmission at high SNR) do not necessarily perform well in terms of mutual information, particularly not in the wideband regime (low spectral efficiency). 
We define the zero-crossing $\nu(\Omega)$ of a constellation $\Omega$ in terms of an SNR measure as the minimum SNR for which $\eta>0$. For the three SNR measures considered in this work ($\gamma_{ E_b}$, $\gamma_{\bar P_o}$, or $\gamma_{\hat P_o}$), the zero-crossings will be denoted as $\nu_{{ E_b}}(\Omega)$, $\nu_{{\bar P_o}}(\Omega)$, and $\nu_{{\hat P_o}}(\Omega)$, respectively.
Fig.~\ref{mi} shows that in terms of $\gamma_{ E_b}$ and $\gamma_{\bar P_o}$, $\mathscr{C}_4$ and OOK have the same lowest zero-crossing of $1.42$ dB and $-0.79$ dB, respectively, among all studied formats. 
However, OOK has the lowest zero-crossing of 2.21 dB in terms of $\gamma_{\hat P_o}$.
The zero-crossings will be explained analytically in this section.
 An interesting question is if there exist other constellations that perform better in the wideband regime than those found in Sec.~\ref{sec:const}. As we shall see in this section, the answer is yes.

\begin{theorem}
For any given $M \geq 2$ with $|\Omega|=M$ and a uniform input distribution,
  \begin{align}
 \label{zc1}   \nu_{{ E_b}}(\Omega) &= 10 \log_{10} \Bigg({ \Bigg(  1- \frac{\|   \sum_{i=0}^{M-1} \mathbf{s}_i  \|^2}{{M}  \sum_{i=0}^{M-1} \|\mathbf{s}_i \|^2}     \Bigg)^{-1} {\log_e 2} }\Bigg),  \\
 \label{zc2}    \nu_{{\bar P_o}}(\Omega) 
&=  5 \log_{10}\Bigg( \frac{(\sum_{i=0}^{M-1} s_{i,1})^2}{M\sum_{i=0}^{M-1} \| \mathbf{s}_i\|^2 - \|  \sum_{i=0}^{M-1} \mathbf{s}_i \|^2}\log_e 2\Bigg) ,
    \end{align}
    \begin{align}
 \label{zc3}  &  \nu_{{\hat P_o}}(\Omega) \nonumber\\
&=  5 \log_{10}\Bigg( \frac{\left(M \max_{i} \left\{s_{i,1} +\sqrt{2(s_{i,2}^2+ s_{i,3}^2)} ~\right\}\right)^2}{M\sum_{i=0}^{M-1} \| \mathbf{s}_i\|^2 - \|  \sum_{i=0}^{M-1} \mathbf{s}_i \|^2}\log_e 2\Bigg) .
    \end{align}
\end{theorem}

\begin{IEEEproof}
According to~\cite[Th.~7]{Agrell}, the zero-crossing 
\begin{equation}\label{ele:zero:crossin}
    \nu_{{ E_b}}(\Omega)= 10 \log_{10} \frac{1}{\alpha(\Omega)}~~\textrm{[dB]},
\end{equation}
where
\begin{equation}\label{alpha:zerocrossing}
\alpha(\Omega)=  \Bigg( 1- \frac{\| \mathbb{E}[\Omega]  \|^2}{ E_s} \Bigg) \log_2 e,
\end{equation}
and $\mathbb{E}[\Omega]$ is the mean of the constellation. Substituting~(\ref{alpha:zerocrossing}) in~(\ref{ele:zero:crossin}) yields~(\ref{zc1}),
whereas substituting~(\ref{zc1}) in~(\ref{exp:gamma:po}) and~(\ref{exp:gamma::hat:po}) yields~(\ref{zc2}) and~(\ref{zc3}), respectively. 
\end{IEEEproof}

The problem of optimizing modulation formats which require the least SNR for reliable communications can be formulated as finding the constellation $\Omega$ which provide the minimum zero-crossing $\nu(\Omega)$. 
It can be represented mathematically as
\begin{eqnarray}
 \text{Minimize~} & & \nu(\Omega)  \\
 \text{Subject to}& & |\Omega| =M\\
   & & \Omega \subset \Upsilon,
\end{eqnarray}
for the three SNR measures considered in this work. Observe that no minimum-distance condition applies.
Using~\cite[Th.~7]{Agrell}, this optimization problem can be solved analytically.

\begin{theorem}\label{zerocrossing:theorem:eb}
For any given $M\geq2$ and $\Omega \subset \Upsilon$ with $|\Omega|=M$,
\begin{equation}
\nu_{{ E_b}}(\Omega) \geq 10 \log_{10}  \frac{M \log_e 2}{ M-1 }  ~~\textrm{[dB]},
\end{equation}
with equality if and only if $M-1$ constellation points are at the origin.
\end{theorem}

\begin{IEEEproof}
%
The numerator of~(\ref{zc1}) can be bounded using
\begin{align}
{ \left|\left|  \sum_{i=0}^{M-1} \mathbf{s}_i \right|\right|^2} &=
 \sum_{i=0}^{M-1} \| \mathbf{s}_i\|^2 + {2 \sum_{i=0}^{M-2} \sum_{j=i+1}^{M-1} \left\langle \mathbf{s}_i, \mathbf{s}_j \right\rangle}    \\
 \label{ineq2ofproof}
 & \geq \sum_{i=0}^{M-1} \| \mathbf{s}_i\|^2,
\end{align}
where $\left\langle \cdot,\cdot \right\rangle$ for $\mathbf{s}_i,\mathbf{s}_j \in \Upsilon$, denotes the inner product, and~(\ref{ineq2ofproof}) is due to the fact that \[
\left\langle \mathbf{s}_i , \mathbf{s}_j \right\rangle = \|  \mathbf{s}_i \| \|  \mathbf{s}_j\| \cos \theta \geq 0, \]
 $\forall i,j=0,\ldots,M-1$, since from Th.~\ref{theorem:cone:admi}, $\max \theta = \cos^{-1}(1/3) =70.528^{\circ} \leq 90^{\circ}$.
 Equality holds if and only if $M-1$ points are located at the origin.
 Applying~(\ref{ineq2ofproof}) in~(\ref{zc1}) completes the proof.
\end{IEEEproof}


Several constellations fulfill Th.~\ref{zerocrossing:theorem:eb}. If the nonzero constellation point has the coordinates $(\sqrt{M E_s},0,0)$, we refer to this constellation as $\mathscr{E}_M$. At the same symbol rate, $\mathscr{E}_M$ occupies a bandwidth of $W=R_s$, as OOK and $M$-PAM.
However, if the nonzero constellation point is at the surface of $\Upsilon$, this point has coordinates $(\sqrt{(2/3)M E_s} , 0 , \sqrt{M E_s/3})$, and will be referred to as $\mathscr{O}_M$. 
The constellation $\mathscr{O}_M$ belongs to the SCM family; therefore it occupies a bandwidth of $W=2R_s$, which is twice the bandwidth of OOK and $M$-PAM at the same symbol rate. 
The reason for choosing the two extremes $\mathscr{E}_M$ and $\mathscr{O}_M$ is that both have the same $\bar P_e$; however, $\mathscr{O}_M$ has lower $\bar P_o$ (see Fig.~\ref{contour}) and $\mathscr{E}_M$ has lower bandwidth.

\begin{theorem}
For any given $M\geq2$ and $\Omega \subset \Upsilon$ with $|\Omega|=M$,
\begin{equation}\label{ave:po:zerocrossingtheorem}
\nu_{{\bar P_o}}(\Omega) \geq 5 \log_{10} \Bigg(  \frac{2}{3(M-1)} \log_e 2    \Bigg)  ~~\textrm{[dB]},
\end{equation}
with equality if and only if $\Omega=\mathscr{O}_M$.

\end{theorem}

\begin{IEEEproof}
By using~(\ref{zc2}), $\nu_{{\bar P_o}}(\Omega)$ can be bounded by
\begin{align}
%
%
\label{firstrelax:ave}
\nu_{{\bar P_o}}(\Omega) &\geq  5 \log_{10} \Bigg( \frac{  \sum_{i=0}^{M-1} s_{i,1}^2  }       {M\sum_{i=0}^{M-1} \| \mathbf{s}_i\|^2 - \|  \sum_{i=0}^{M-1} \mathbf{s}_i \|^2}\log_e 2\Bigg) \\
\label{secondrelax:ave}
&\geq  5 \log_{10} \Bigg( \frac{  \sum_{i=0}^{M-1} s_{i,1}^2  }       {(M-1)\sum_{i=0}^{M-1} \| \mathbf{s}_i\|^2 }\log_e 2\Bigg) \\
\label{thirdrelax:ave}
&\geq  5 \log_{10} \Bigg( \frac{   2 }       {3(M-1) }\log_e 2\Bigg). 
\end{align} 
 The inequality in
 (\ref{firstrelax:ave}) follows from $s_{i,1}, s_{j,1} \geq 0, ~\forall i,j=0,1,\dots,M-1$, whereas
(\ref{secondrelax:ave}) follows from~(\ref{ineq2ofproof}) and
(\ref{thirdrelax:ave}) follows from the definition of the admissible region in~(\ref{adm:scm}).
This completes the proof of~(\ref{ave:po:zerocrossingtheorem}). 

To prove that~(\ref{ave:po:zerocrossingtheorem}) is tight if and only if $\Omega =\mathscr{O}_M$, we observe that~(\ref{firstrelax:ave}) and~(\ref{secondrelax:ave}) are tight if and only if $M-1$ points are at the origin, while~(\ref{thirdrelax:ave}) is tight if and only if all $M$ points are located on the boundary of the cone.

%
%
\end{IEEEproof}

So far, we were able to derive optimal constellations in terms of $\nu_{E_b}$ and $\nu_{\bar P_o}$. For the third SNR measure $\nu_{\hat P_o}$, however, we resort to a conjecture only, for which we have solid numerical support but no proof. 
\begin{conjecture}

For any $\Omega \subset \Upsilon$,
\begin{equation}
\nu_{\hat P_o} \geq 5 \log_{10}(4 \log_e 2),
\end{equation} 
with equality if and only if $\Omega=\mathscr{E}_2$ (OOK).
\end{conjecture}

\begin{figure}
\centering
\footnotesize
\psfrag{aaaaaaa1}[c][c][0.75][0]{OOK}
 \psfrag{aaaaaaa2}[c][c][0.75][0]{QPSK }
\psfrag{aaaaaaa3}[c][c][0.9][0]{$\mathscr{C}_4$ }
\psfrag{aaaaaaa4}[c][c][0.75][0]{8-QAM}
 \psfrag{aaaaaaa5}[c][c][0.75][0]{$\mathscr{L}_{\bar P_e,8}$}
\psfrag{aaaaaaa6}[c][c][0.9][0]{$\mathscr{C}_{\bar P_e,8}$}
\psfrag{aaaaaaa7}[c][c][0.9][0]{$\mathscr{C}_{\bar P_o,8}$}
\psfrag{aaaaaaa8}[c][c][0.75][0]{$\mathscr{C}_{\hat P_o,8}$}
\psfrag{aaaaaaa9}[c][c][0.75][0]{4-PAM}
\psfrag{aaaaaaa10}[c][c][0.75][0]{16-QAM}
\psfrag{aaaaaaa11}[c][c][0.75][0]{$\mathscr{L}_{16}$}
\psfrag{aaaaaaa12}[c][c][0.9][0]{$\mathscr{C}_{\bar P_e,16}$}
\psfrag{aaaaaaa13}[c][c][0.9][0]{$\mathscr{C}_{\bar P_o,16}$}
\psfrag{aaaaaaa14}[c][c][0.9][0]{$\mathscr{C}_{\hat P_o,16}$}
\psfrag{xxxxxxx4}[c][c][0.75][0]{$\breve 8$-QAM}
\psfrag{xxxxxxx5}[c][c][0.75][0]{8-PSK}
\psfrag{xxxxxxx10}[c][c][0.75][0]{$\breve {16}$-QAM}
\psfrag{aaaaaaa18}[c][c][0.75][0]{16-PSK}
\psfrag{sssssss101}[c][c][0.75][0]{$\mathscr{E}_4$}
\psfrag{sssssss102}[c][c][0.75][0]{$\mathscr{E}_8$}
\psfrag{sssssss103}[c][c][0.75][0]{$\mathscr{E}_{16}$}

\psfrag{sssssss201}[c][c][0.75][0]{$\mathscr{O}_4$}
\psfrag{sssssss202}[c][c][0.75][0]{$\mathscr{O}_8$}
\psfrag{sssssss203}[c][c][0.75][0]{$\mathscr{O}_{16}$}

\psfrag{ebnodb}[t][c][1][0]{$\gamma_{ E_b}~\text{[dB]}$}
\psfrag{MI}[c][c][1][0]{$\eta$ [bit/s/Hz]}
\psfrag{TbavePosquaredNodB}[t][c][1][0]{$\gamma_{\bar P_o}~\text{[dB]}$}
\psfrag{TbpeakPosquaredNodB}[t][c][1][0]{$\gamma_{\hat P_o}~\text{[dB]}$}

\begin{tabular}{c}
\includegraphics[scale=0.51]{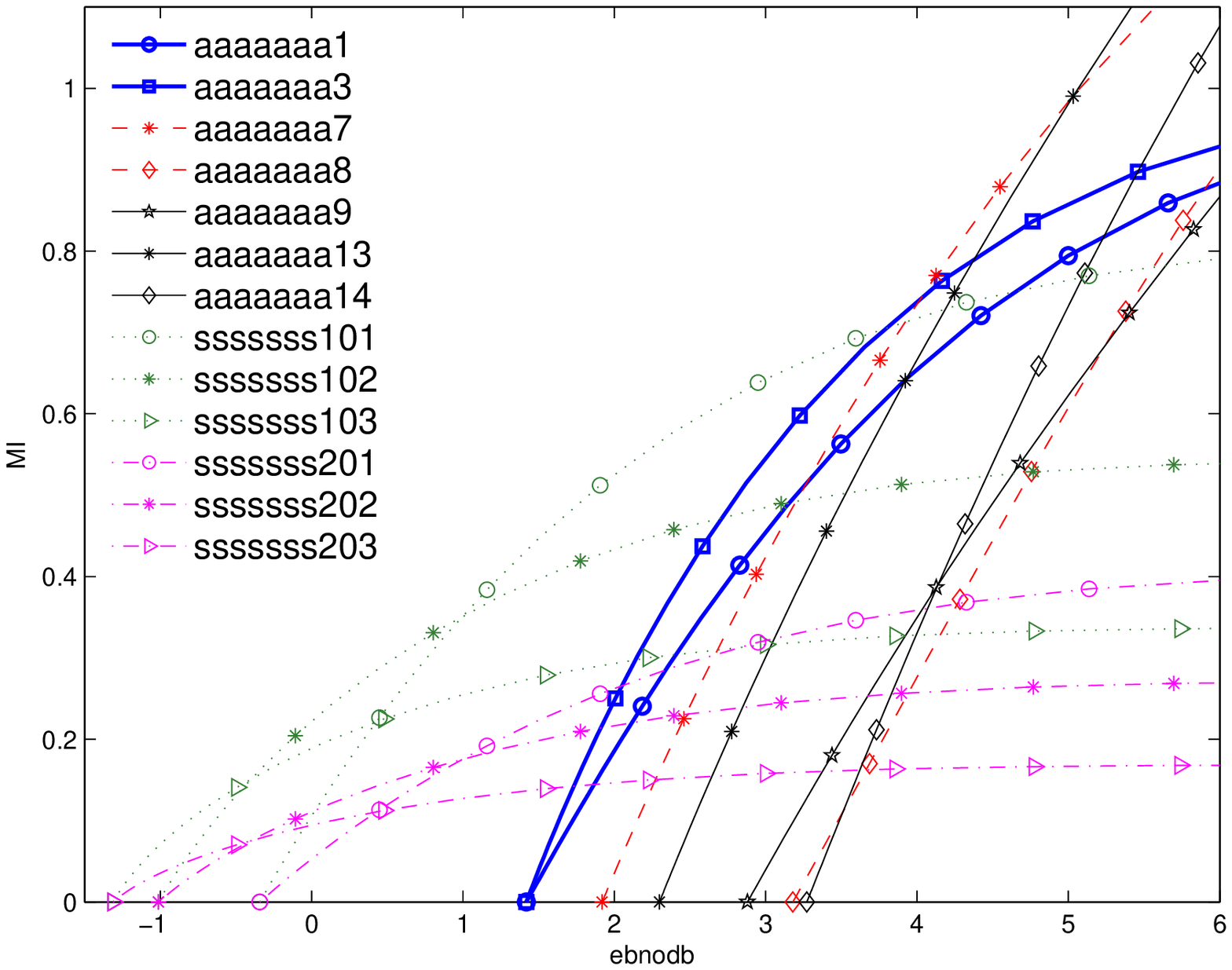} \\ \\
\includegraphics[scale=0.51]{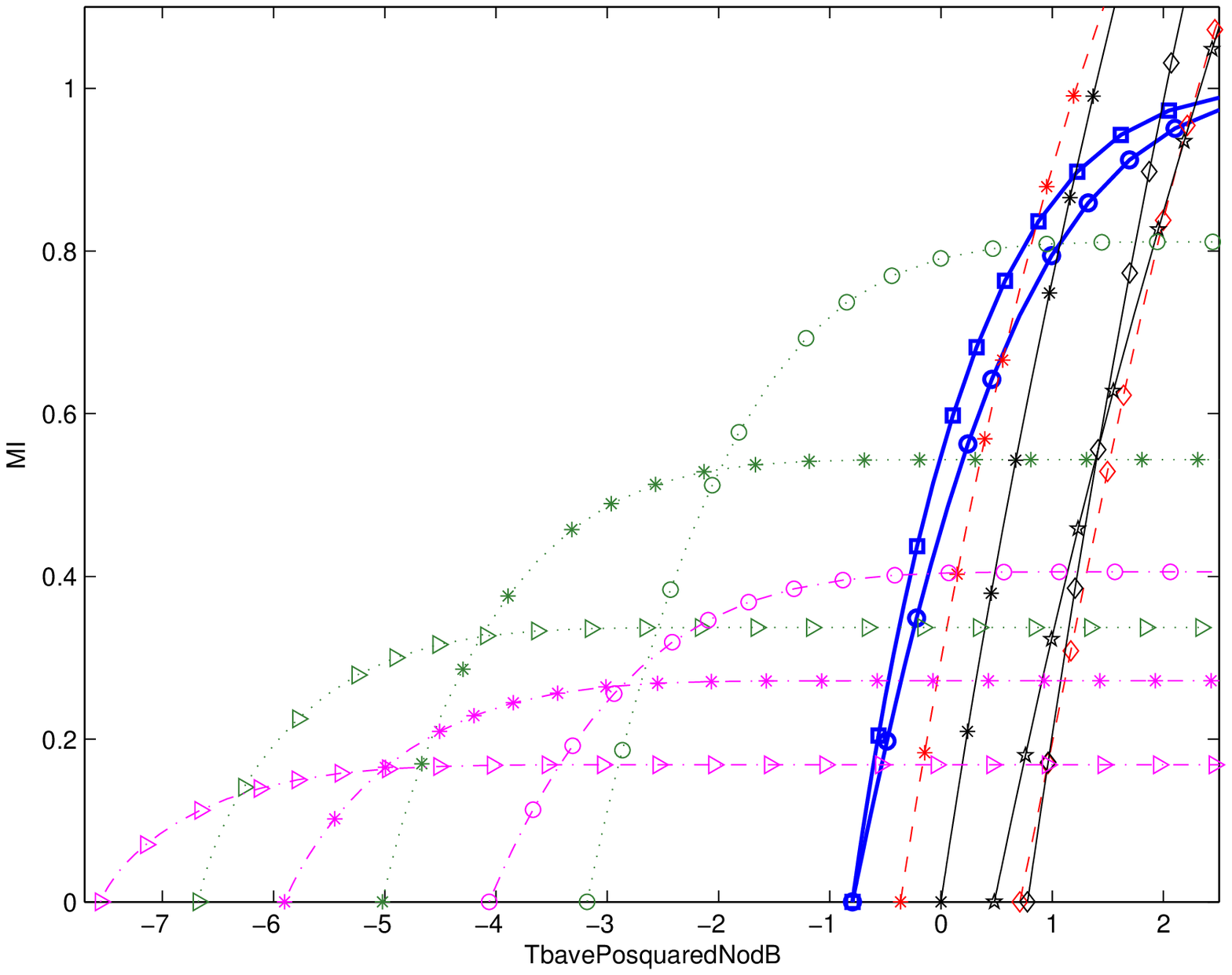} \\ \\
\includegraphics[scale=0.51]{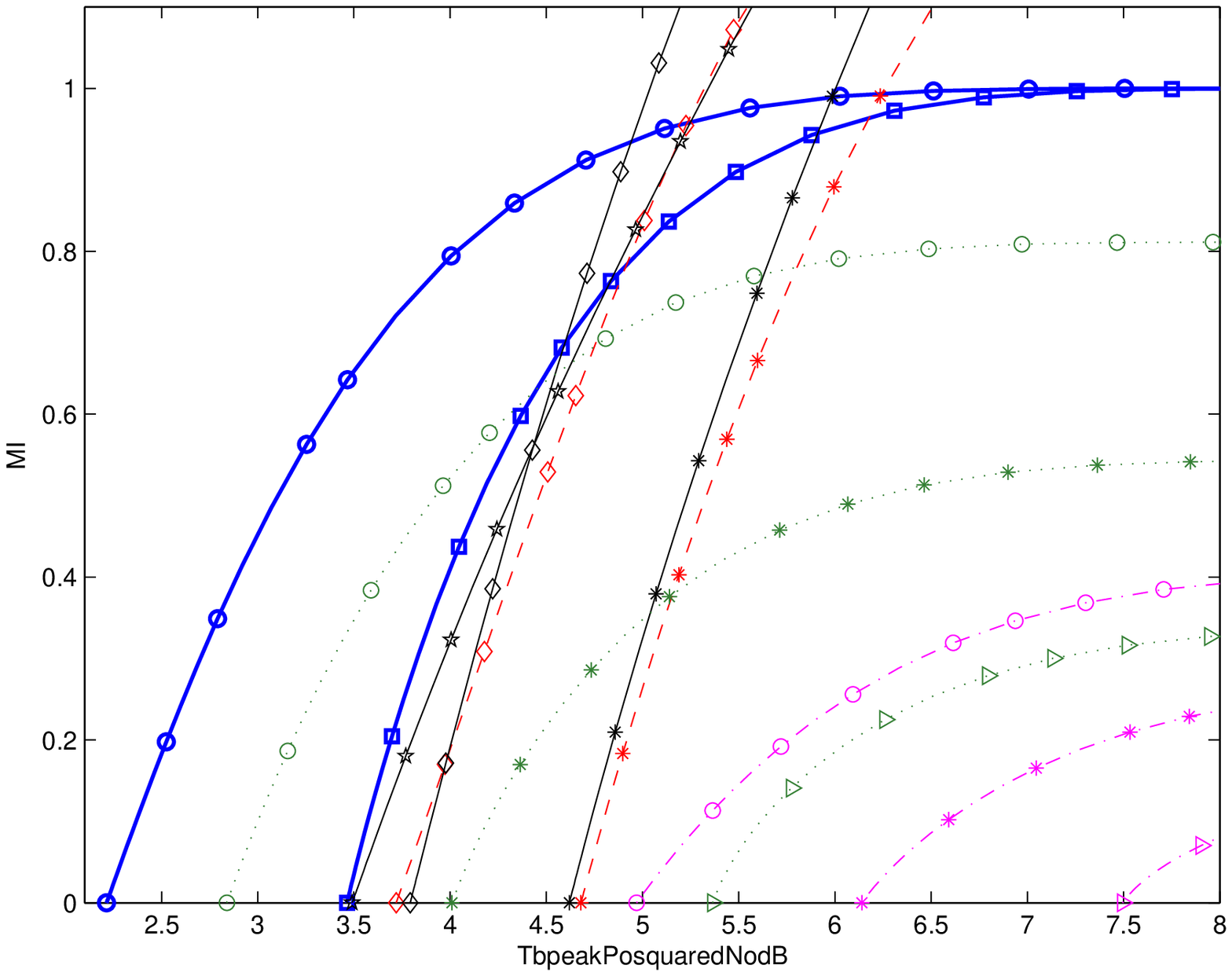} \\ \\
\end{tabular}
\caption{Spectral efficiency of the optimal constellations $\mathscr{E}_M$ and $\mathscr{O}_M$ in the wideband regime. Some of the previously optimized constellations are included for comparison.}
\label{mizoom}
\end{figure}

Fig.~\ref{mizoom} (top) shows the spectral efficiency in the low $\gamma_{E_b}$ regime
for $\mathscr{E}_M$ and $\mathscr{O}_M$, where $M=4$, $8$, and $16$. We also include some of the previously optimized constellations for comparison.
Even though $\mathscr{E}_{M}$ and $\mathscr{O}_{M}$, for a given $M$, have the same zero-crossing $\nu_{E_b}$, $\mathscr{E}_{M}$ performs better at all nonzero spectral efficiencies. Indeed, $\mathscr{E}_{16}$, $\mathscr{E}_{8}$, and $\mathscr{E}_{4}$ are the best of all studied constellations for $\eta <0.70$.
%
As $M \rightarrow \infty$, $\mathscr{E}_{M}$ and $\mathscr{O}_{M}$ approach the coherent (non-IM/DD) Shannon limit, i.e., $\nu_{E_b}(\mathscr{E}_{M})= \nu_{E_b}(\mathscr{O}_{M}) \rightarrow -1.59$ dB.


Fig.~\ref{mizoom} (middle) depicts the spectral efficiency in the low $\gamma_{\bar P_o}$ regime. For $M=4$, $8$, and $16$, the formats $\mathscr{O}_{M}$ have lower $\nu_{{\bar P_o}}(\Omega)$ than $\mathscr{E}_{M}$; however, the latter performs better than all other studied formats for $0.13< \eta <0.80$. Using~(\ref{ave:po:zerocrossingtheorem}), $ \nu_{{\bar P_o}}(\Omega)$ approaches $ -\infty$ as $M \rightarrow \infty$ faster for $\Omega=\mathscr{O}_{M}$ than $\mathscr{E}_{M}$.


Finally, we show the spectral efficiency in the low $\gamma_{\hat P_o}$ regime in Fig.~\ref{mizoom} (bottom). Surprisingly, OOK has the best performance up to $\eta=0.93$,  and then $\mathscr{C}_{\hat P_o,16}$ outperforms the rest. 
On the other hand, the modulation formats which minimize $\nu_{E_b}(\Omega)$ or $\nu_{{\bar P_o}}(\Omega)$ do not have the best performance as before. The trend is that their performance is poorer with increased number of levels.
Also, $\mathscr{O}_{M}$ is worse than $\mathscr{E}_{M}$ for a given $M$, since the former has a higher peak power requirement due to the nonzero point on the surface of $\Upsilon$.

It may seem somewhat unexpected that a constellation with $M-1$ points at the origin would perform better at low SNR than all other constellations. 
Such a constellation is equivalent to a binary constellation with a higher probability for the zero point. 
%
Similar constellations were found to achieve capacity in~\cite{Channels2005} and~\cite{Smith1971}, albeit for different channels.
A similar constellation was studied in~\cite{Steiner1994}, where a signal set for coherent AWGN channels consisting of two antipodal signals and $M-2$ signals at the origin was used to disprove the strong simplex conjecture, according to which the regular simplex signal set would minimize the uncoded probability of error under an average energy constraint. 


\section{Conclusions}\label{sec:conclusion}
By using the minimum distance as a modulation design criterion for uncoded systems, we were able to numerically optimize 4-, 8-, and 16-level single-subcarrier IM/DD modulation formats for systems which are limited by average electrical, average optical, and peak power.
For $M=4$, the most power-efficient modulation in terms of average electrical, optical, and peak power has a tetrahedral structure. This constellation is also a subset of all the obtained higher-level constellations. 
As for the 8- and 16-level constellations, power-efficient schemes are obtained by not confining the set of constellation points to a regular structure such as that of a lattice. 
However, this comes at the price of losing the geometric regularity, which increases the modulator and demodulator complexity. 
%
Our comparisons show that the penalty gap for using modulation formats optimized for $\hat P_o$ in systems which are $\bar P_e$ or $\bar P_o$ limited is much less than the penalty gap when using formats optimized for $\bar P_e$ or $\bar P_o$ in systems which are limited by $\hat P_o$.
Therefore, modulation formats optimized for $\hat P_o$ should be preferred in applications with mixed power requirements.
The overall gain of the obtained formats over the previously best known formats is between 0.6 dB and 3 dB at asymptotically high SNR. We conjecture that the new obtained modulation formats are optimal for their size and optimization criteria over uncoded single-subcarrier IM/DD channels.

On the other hand, when capacity-achieving error-correcting codes are deployed,
the best modulation formats in the wideband regime (small $\eta$) have only one nonzero constellation point. This is confirmed analytically and numerically.
At higher $\eta$, modulation formats optimized for $\bar P_o$ are able to achieve higher reliable transmission rates compared to other formats in systems which are limited by average electrical or optical power. 
OOK and the 16-level modulation format optimized for peak power offer the best performance in peak-power limited systems.
The overall gain of the obtained formats over previously best known formats ranges between 0.3 dB and 1 dB.

\appendices
\section{Obtained Constellations}\label{bestconstapp}
This appendix lists the coordinates of the numerically optimized constellations in Sec.~\ref{subsec:optimconst}. Whenever possible, numerical values have been replaced with the corresponding exact values. Constellations are normalized to unit $d_{\text{min}}$.
\begin{align*}
 \mathscr{C}_{4}&=\mathscr{C}_{\bar P_e,4}=  \mathscr{C}_{\bar P_o,4}= \mathscr{C}_{\hat P_o,4}=\mathscr{L}_{\bar P_e,4}=\mathscr{L}_{\bar P_o,4} =\mathscr{L}_{\hat P_o,4}=\\
 &\{(0 , 0, 0),
(\sqrt{2/3} , 0 , 1/\sqrt{3}),
(\sqrt{2/3},  \pm 1/2 , -\sqrt{3}/6)\}.\\ \\
%
%
 \mathscr{C}_{\bar P_e,8}&=  \mathscr{C}_4 \cup \{
 ((5/3)\sqrt{2/3}  ,0,  -5/(3\sqrt{3})),\\
&((5/3)\sqrt{2/3} ,\pm 5/6, 5/(6\sqrt{3})),
(2 \sqrt{2/3}   ,                0  ,                 0)
\}. \\ \\
%
%
\mathscr{C}_{\bar P_o,8}&=  \mathscr{C}_4 \cup \{
((5/3)\sqrt{2/3},  0, -5/(3\sqrt{3})),\\
&((5/3)\sqrt{2/3},   \pm 5/6 ,  5/(6\sqrt{3})),\\
&(1.6293   ,-0.9236,  -0.6886)
\}. \\ \\
%
%
\mathscr{C}_{\hat P_o,8}&=  \mathscr{L}_{\hat P_o,8}=\mathscr{C}_4 \cup \{
(2 \sqrt{2/3} ,  0  ,-1/ \sqrt{3}),\\
&(2 \sqrt{2/3} , \pm 1/2   ,\sqrt{3}/6),
(\sqrt{6}   ,       0 ,       0)
\}. \\ \\
%
%
 \mathscr{L}_{\bar P_e,8}&=  \mathscr{L}_{\bar P_o,8}=\mathscr{C}_4 \cup \{
 (2\sqrt{2/3}, \pm1/2, \sqrt{3}/6),\\
&(2\sqrt{2/3},         0, -1/\sqrt{3}),
(        2\sqrt{2/3}, 1, -1/\sqrt{3})
\}. \\ \\
%
%
%
%
\mathscr{C}_{\bar P_e,16}&=  \mathscr{C}_4 \cup \{ (1.3608   ,5/6  , 5/(6\sqrt{3})), \\
&(1.3608,  0  ,-0.9623), (1.4628 , -0.7513   ,0.7110),\\
&(1.6024  ,-1.1134,  -0.2106), (1.6293   ,0.1346 ,  1.1442),\\
&(1.6293,   0.9236  ,-0.6887), (2 \sqrt{2/3} ,                  0   ,                0),\\
&(1.9336  ,-0.8075,  -1.1032), (2.0380   ,1.4396 ,  0.0642),\\
&(2.3097 ,  0.5202  , 0.5210), (2.3097  , 0.1911  ,-0.7110),\\
&(2.3499  ,-0.6462  , 0.2616)
\}. \\ \\
\end{align*}
\begin{align*}
\mathscr{C}_{\bar P_o,16} &= \mathscr{C}_{\bar P_o,8} \cup \{ (1.6293,  -0.1345,  1.1442),\\
&(1.6293 ,  1.0582 , -0.4556), (2 \sqrt{2/3}   ,                0 ,                  0),\\
&(2.0380,   0.6643 , -1.2789), (2.0380 , -1.4396 ,  0.0642),\\
&(2.0380  , 0.7754 ,  1.2147), (2.1187   ,1.4645 ,  0.3160),\\
&(2.1187 , -1.0059 ,  1.1103)
\}. \\ \\ 
%
%
%
%
\mathscr{C}_{\hat P_o,16}&= \mathscr{C}_4 \cup \{ (1.6279,   0.8995,  -0.7184),\\
&(1.6279 , -0.4977 ,  1.0379), \\
&(1.6270  ,-0.9003  ,-0.7162), (1.6300   ,0.5022,   1.0374),\\
& (1.6310,  -0.0010 , -1.1533), \\
&(1.6313 , -1.1242  , 0.2584), (1.6328  , 1.1259   ,0.2557), \\
& (2 \sqrt{2/3}   ,                0 ,                  0), (2.4495,   0 , 1/\sqrt{3}),\\
& (2.4495 , \pm 1/2 , -\sqrt{3}/6), (3.2660   ,                0   ,                0)
\}. \\ \\ 
%
%
%
%
%
%
%
\mathscr{L}_{16}&=\mathscr{L}_{\bar P_e,16}=\mathscr{L}_{\bar P_o,16} =\mathscr{L}_{\hat P_o,16}= \mathscr{L}_{\bar P_e,8}  \cup \{ \\
&2\sqrt{2/3}, 0, (2/3)\sqrt{3}),\\
&(      2\sqrt{2/3}, - 1, -\sqrt{3}/3),
(        \sqrt{6}, 0, 0 ),\\
&(      \sqrt{6}, -1/2, \sqrt{3}/2),
(      \sqrt{6}, \pm 1/2, -\sqrt{3}/2),\\
&(    \sqrt{6}, \pm 1, 0 )
\}.
\end{align*}

\section*{Acknowledgment}

The authors would like to acknowledge LINDO Systems for the free license to use their numerical optimization software.
We would also like to acknowledge Dr. Giuseppe Durisi and Dr. Alex Alvarado for the interesting discussions about mutual information, Tilak Rajesh Lakshmana for the nice discussions about optimization techniques, and Rajet Krishnan for his comments about the paper structure.

\bb{

\ifCLASSOPTIONcaptionsoff
  \newpage
\fi

}
\end{document}